\documentclass{amsproc}
\usepackage{amssymb,graphics}

\newtheorem{theorem}{Theorem}[section]
\newtheorem{proposition}[theorem]{Proposition}
\newtheorem{corollary}[theorem]{Corollary}

\theoremstyle{definition}

\theoremstyle{remark}
\newtheorem{remark}[theorem]{Remark}

\numberwithin{equation}{section}

\def\bbblb#1{\setbox\@tempboxa\hbox{$#1[$}%
             \@tempdimb\wd\@tempboxa
             \copy\@tempboxa \kern -.85\@tempdimb
             \copy\@tempboxa \kern -.65\@tempdimb\box\@tempboxa}
\def\bbbrb#1{\setbox\@tempboxa\hbox{$#1]$}%
             \@tempdimb\wd\@tempboxa
             \copy\@tempboxa \kern -.65\@tempdimb
             \copy\@tempboxa \kern -.85\@tempdimb\box\@tempboxa}

\def\cal#1{\frak #1}
\def\openone{\leavevmode\hbox{\small1\kern-3.3pt\normalsize1}}
\def\bbbone{\leavevmode\hbox{\small1\kern-3.3pt\normalsize1}}

\def\opl{\mathop{\oplus\,}\limits}
\def\bigla{\big\langle\hspace*{-.9mm}\big\langle}
\def\bigra{\big\rangle\hspace*{-.9mm}\big\rangle}
\def\Bigla{\Big\langle\hspace*{-1.0mm}\Big\langle}
\def\Bigra{\Big\rangle\hspace*{-1.0mm}\Big\rangle}
\def\otimescomma{\mathop{\otimes}\limits_{'}}

\def\diag{\text{diag\,}}
\def\res{\mathop{{\rm Res}\,}\limits}
\def\lnu{\mathop{{<}\,}\limits_\nu }
\def\gnu{\mathop{{>}\,}\limits_\nu }
\def\genu{\mathop{{\geq}\,}\limits_\nu }
\def\lenu{\mathop{{\leq}\,}\limits_\nu }
\def\ad{\text{ad\,}}
\def\tr{\text{tr\,}}
\def\im{\text{Im\,}}
\def\cotan{\text{cotan\,}}

\def\fr#1{{\frak #1}}
\def\bbbr{\Bbb{R}}
\def\bbbz{\Bbb{Z}}

\def\bbbc{\Bbb{C}}

\allowdisplaybreaks

\begin{document}

\title{Algebraic and Analytic Aspects of  Soliton Type Equations}

\author{Vladimir S. Gerdjikov}
\address{Institute for Nuclear Research and Nuclear Energy,
Tzarigradsko chaussee 72, 1784 Sofia, Bulgaria}
\curraddr{Institute for Nuclear Research and Nuclear Energy,
Tzarigradsko chaussee 72, 1784 Sofia, Bulgaria}
\email{gerjikov@inrne.bas.bg}
\thanks{Financial support in part from  Gruppo collegato di INFN at
Salerno, Italy and project PRIN 2000 (contract 323/2002) is acknowledged.}

%    General info
\date{}



\begin{abstract}
This is a review of two of the fundamental tools for analysis of
soliton equations: i) the algebraic ones based on Kac-Moody
algebras, their central extensions and their dual algebras which underlie
the Hamiltonian structures of the NLEE; ii) the construction of the
fundamental analytic solutions (FAS) of the Lax operator and the
Riemann-Hilbert problem (RHP) which they satisfy. The fact that
the inverse scattering problem for the Lax operator can be viewed
as a RHP gave rise to the dressing Zakharov-Shabat, one of the
most effective ones for constructing soliton solutions. These two
methods when combined may allow one to prove rigorously the
results obtained by the abstract algebraic methods. They also
allow to derive spectral decompositions for non-self-adjoint Lax
operators.

\end{abstract}

\maketitle %\tableofcontents

\specialsection{Introduction}
%\section{Introduction}
%This is an example of an unnumbered first-level heading.

%\section*{Lax representations}\label{sec:In.1}

We start with three examples of integrable nonlinear evolution equations
(NLEE). The first one is the well known $N $-wave equation
\cite{ZaMa,ZMNP,Kaup}:
\begin{equation}\label{eq:3.1}
i[I,Q_t] - i [J,Q_x] + [[I,Q],[J,Q(x,t)]]=0,  \qquad
\lim_{x\to\pm\infty } Q(x,t) =0,
\end{equation}
where $Q(x,t)$ is a smooth $n\times n $ matrix-valued
function, $Q(x,t)=-BQ^\dag B $ and $I $ and $J $ are constant
diagonal matrices; $B_{ij}=\delta _{ij}\epsilon _j$, $\epsilon _j=\pm 1$.

The second example is the 2-dimensional affine Toda chain \cite{Mikh}:
\begin{equation}\label{eq:3.2}
{\partial^2 Q_k \over \partial x \partial t } = e^{Q_{k+1}-Q_{k}} -
e^{Q_{k}-Q_{k-1}}, \qquad k=1,\dots, n,
\end{equation}
where we assume that $e^{Q_{n+1}}\equiv e^{Q_1} $.

The third example belongs to the same family as (\ref{eq:3.2}) and is of
the form:
\begin{equation}\label{eq:3.3}
i{\partial\psi_k \over \partial t } + \gamma \coth {\pi k \over n }
{\partial^2\psi _k \over \partial x^2 } + i \gamma \sum_{p=1}^{n-1}
{d  \over dx }\left( \psi _p\psi _{k-p} \right)=0, \qquad k=1,\dots , n,
\end{equation}
and $k-p $ is understood modulo $n $ and $\psi_0=\psi_n=0$.

The integrability of these equations is based on their Lax
representations. This means that each of the NLEE
(\ref{eq:3.1})--(\ref{eq:3.3}) can be represented as the compatibility
condition
\begin{equation}\label{eq:lax}
[L(\lambda ),M(\lambda )]=0,
\end{equation}
of two linear matrix differential operators depending on the spectral
parameter $\lambda $. Below we will use as Lax operator $L(\lambda ) $
\begin{equation}\label{eq:2.2}
L(\lambda )\psi (x,t,\lambda ) = \left( i {d \over dx } + q(x,t) -
\lambda J \right) \psi (x,t,\lambda ) =0;
\end{equation}
as examples of $M(\lambda ) $-operators we use:
\begin{subequations}\label{eq:2.3}
\begin{eqnarray}\label{eq:2.3a}
&&\qquad  M(\lambda )\psi  = \left( i {d \over dt } + V_0(x,t) +
\lambda I \right) \psi (x,t,\lambda ) =
\lambda \psi (x,t,\lambda ) I ; \\
\label{eq:2.3b}
&&\qquad  M_1(\lambda )\psi = \left( i {d \over dt } + V_0(x,t) +
\lambda V_1(x,t) + \lambda ^2 V_2 \right) \psi (x,t,\lambda ) =
\lambda ^2 \psi (x,t,\lambda ) V_2^{\rm as}; \\
\label{eq:2.3c}
&& \qquad  M_2(\lambda )\psi  = \left( i {d \over dt } + V_0(x,t) +
{1 \over \lambda} V_{-1}(x,t) \right) \psi (x,t,\lambda ) =
{1  \over \lambda  } \psi (x,t,\lambda ) V_{-1}^{\rm as};
\end{eqnarray}
\end{subequations}
where $V_{2}^{\rm as} =\lim_{x\to\pm\infty } V_2(x,t) $ and
$V_{-1}^{\rm as} =\lim_{x\to\pm\infty } V_{-1}(x,t) $.

Choosing the form of $L(\lambda ) $ in (\ref{eq:2.2}) we fixed up the
gauge by assuming that $J $ is constant diagonal matrix and $q(x,t)=
[J,Q(x,t)] $, i.e. $q_{jj}=0 $.

The system (\ref{eq:2.2}) with $q(x,t) $ and $J $ $2\times 2
$-matrices (i.e., $\fr{g}\simeq sl(2) $) is known as the
Zakharov-Shabat (ZSs) system; the same system with $n\times n $
matrices will be referred to below as the generalized Zakharov-Shabat
system (GZSs).

The Lax representation of the $N $-wave equation is provided by
$L(\lambda )$ (\ref{eq:2.2}) and $M(\lambda )$ (\ref{eq:2.3a}).
If the potentials in these operators are $n\times n $-matrix ones we may
assume that the Lie algebra underlying the Lax representation is
$\fr{g}\simeq sl(n) $. The set of independent fields $Q_{ij}(x,t) $ equals
$n(n-1) $ and may be restricted by the involution \cite{ZaMa,ZaSha,ZMNP}:
\begin{equation}\label{eq:3.4}
q(x,t) = B q^\dag (x,t) B^{-1}, \qquad J = J^\dag,
\end{equation}
Often by $N $-wave equations in the
literature people mean eq. (\ref{eq:3.1}) with the involution
(\ref{eq:3.4}). Such systems with $n=3 $ and $n=4 $ find applications in
describing wave-wave interactions, see \cite{ZMNP,ZaMa,Kaup}.

The Lax representation of the $\bbbz_n $-NLS eq. (\ref{eq:3.3}) is
provided by (\ref{eq:2.2}) and (\ref{eq:2.3b})  but with rather specific
restrictions imposed on $q(x,t) $ and $J$:
\begin{equation}\label{eq:4.2}
\tilde{q}(x,t) = i\sum_{k=1}^{n} \psi _k(x,t) K_{0}^k, \qquad \tilde{J}
=c_0\sum_{k=1}^{n} \omega ^{k-1/2}E_{kk}, \qquad K_0=\sum_{k=1}^{n}
E_{k,k+1},
\end{equation}
Here and below we will denote by $E_{jk} $ the $n\times n $-matrices equal
to $(E_{jk})_{mn} =\delta _{jm}\delta _{kn} $; the indices should be
taken modulo $n $, i.e. $E_{n,n+1}\equiv E_{n,1} $ and the constant $c_0
$ will be defined below.

The affine Toda chain (\ref{eq:3.2}) has several equivalent Lax
representations. We mention here two of them. Their Lax operators are:
\begin{equation}\label{eq:To1}
\tilde{L}_{\rm Toda} = i {d  \over dx } - {i  \over 2 } \sum_{k=1}^{n}
{dQ_k  \over dx } E_{kk} + i \lambda  \sum_{k=1}^{n} e^{(Q_{k+1}-Q_k)/2}
E_{k,k+1},
\end{equation}
and its gauge equivalent:
\begin{equation}\label{eq:To2}
\tilde{\tilde{L}}_{\rm Toda} = i {d  \over dx } - i \sum_{k=1}^{n} {dQ_k
\over dx } E_{kk} + i \lambda K_0 .
\end{equation}
The corresponding $M $-operators are of the form (\ref{eq:2.3c}). Both
choices (\ref{eq:To1}) and (\ref{eq:To2}) are not of the form
(\ref{eq:2.2}), but are adjusted to the grading of the Lie algebra
$sl(n,\bbbc) $ we introduce in the next subsection, see formulae
(\ref{eq:5.1})--(\ref{eq:7.2}) below.

The operator $\tilde{\tilde{L}}_{\rm Toda} $ (\ref{eq:To2}) after a
similarity transformation with the constant matrix $u_0 $, such that
$u_0^{-1} K_0 u_0 = \sum_{k=1}^{n}\omega ^kE_{kk} $ can be cast into the
form of (\ref{eq:2.2}) in which both $q(x,t) $ and $J $ have a special
form:
\begin{equation}\label{eq:To3}
\tilde{q}(x,t) = -i \sum_{k=1}^{n} {dQ_k  \over dx } \omega ^{kp} K_0^p,
\qquad \tilde{J}=  c_0\sum_{j=1}^{n} \omega ^{k-1/2} E_{kk}.
\end{equation}
where $\omega =\exp (2\pi i /n ) $.  The special form of $q (x,t)$ and $J
$ in both (\ref{eq:4.2}) and (\ref{eq:To3})  shows that both models have
only $n-1 $ independent fields. This special form can also be made
compatible with the structure of the graded and Kac-Moody algebras
\cite{DrinSok,Kac,Helg} and is best understood with the method of
the reduction group proposed by Mikhailov \cite{Mikh}.

The idea of the ISM is based on the possibility to linearize the NLEE
\cite{ZaSh*72,AKNS,CalDeg,FaTa,ZaMa,ZMNP,Kaup}.
To this end we consider the solution to the NLEE $q(x,t) $ as a potential
in $L(\lambda ) $ (\ref{eq:2.2}). In order to solve the
direct scattering problem for $L(\lambda ) $ we introduce the Jost
solutions $\psi _\pm(x,t,\lambda ) $ and the scattering matrix
$T(\lambda,t) $ as follows:
\begin{eqnarray}\label{eq:10.1}
&& i{d\psi _\pm  \over dx } +(q(x,t) -\lambda J) \psi _\pm(x,t,\lambda )
=0,\\
\label{eq:10.2}
&& \lim_{x\to\pm\infty } \psi _\pm(x,t,\lambda ) e^{i\lambda Jx} =\openone
,\\
\label{eq:10.3}
&& T(\lambda ,t) = \psi _+^{-1}\psi _-(x,t,\lambda ).
\end{eqnarray}

The Jost solutions of $L(\lambda ) $ are also eigenfunctions of the
operator $M(\lambda ) $.  We can use this fact to determine the $t
$-dependence of the scattering matrix:
\begin{equation}\label{eq:10.4}
i{dT  \over dt } + [f(\lambda ),T(\lambda ,t)] =0,
\end{equation}
which can be easily solved as follows:
\begin{equation}\label{eq:10.4'}
T(\lambda ,t) = e^{if(\lambda )t} T(\lambda ,0) e^{-if(\lambda )t}.
\end{equation}
By $f(\lambda )\in \fr{h} $ above we mean the dispersion law of
the NLEE; for the $N $ -wave system we have $f_{\rm
N-w}(\lambda)=\lambda I $.

Thus the solution of the NLEE for a given initial condition $q(x,t)|_{t=0}
= q_0(x)$ can be performed in three steps, see \cite{ZMNP,CalDeg,FaTa}:

\begin{enumerate}

\item insert $q(x,0) $ as a potential in $L(\lambda ) $ and determine
the corresponding scattering matrix $T(\lambda ,0) $;

\item Given $T(\lambda ,0) $ and the dispersion law $f(\lambda ) $ find
$T(\lambda ,t) $ from eq. (\ref{eq:10.4'});

\item Given $T(\lambda ,t) $ reconstruct the corresponding potential
$q(x,t) $ of $L(\lambda ) $ which will be also the solution of the NLEE.

\end{enumerate}

Step 2 is trivial. Steps 1 and 3 involve solving the direct and
inverse scattering problem for (\ref{eq:2.2}) which can be reduced
to linear integral equations. The most difficult step 3 provided
for the name of the method. The most effective method to solve it
for operators like $ L(\lambda ) $ is based on the equivalence to
a RHP \cite{Sha}.

Along with solving the inverse scattering problem in step 3) we will
construct also the minimal set of scattering data ${\cal  T} $. Indeed the
scattering matrix $T(\lambda ,t) $ has $n^2 $ matrix elements with only
one obvious constraint $\det T(\lambda ,t)=1 $ while the potential $q(x,t)
$ has only $n(n-1) $ matrix elements. Therefore there must be additional
interrelations between the matrix elements of $T(\lambda ,t) $.

The analysis of the mapping between $q(x,t) $ and ${\cal  T} $ allows one
to interprete it as a generalized Fourier transform
\cite{AKNS,KauNew,GeKh,Ge4,GeKu,GeYa*85,GeYa*86,GeYa}.  The proof of all
these facts and the effective solving of the ISP for the GZSs
(\ref{eq:2.2}) is based on the possibility to construct fundamental
solutions of (\ref{eq:2.2}) which are section-analytic functions of the
spectral parameter $\lambda $.

\section*{Algebraic structures: Kac-Moody and graded Lie
algebras}\label{sec:In.2}

Let us now briefly outline the first basic tool inherent in the
Lax representation -- its algebraic structure. Indeed, $L(\lambda ) $ and
$M(\lambda ) $ above are polynomial in $\lambda  $ and/or $1/\lambda  $
whose coefficients take values in some simple Lie algebra $\fr{g} $.

Let us take generic Lax operators in the form:
\begin{eqnarray}\label{eq:lax-gen}
L(\lambda )\psi\equiv \left(i {d  \over dx } + U(x,t,\lambda ) \right)
\psi (x,t,\lambda )=0, \\
M(\lambda )\psi\equiv \left(i {d  \over dt } + V(x,t,\lambda ) \right)
\psi (x,t,\lambda )=\psi (x,t,\lambda )V^{\rm as}(\lambda), \\
\label{eq:1.18}
U(x,t,\lambda ) = \sum_{k}^{}U_k(x,t) \lambda ^k,  \qquad
V(x,t,\lambda ) = \sum_{k}^{}V_k(x,t) \lambda ^k,
\end{eqnarray}
where the potentials $U(x,t,\lambda ) $ and $V(x,t,\lambda )  $ are
polynomials in $\lambda  $ and/or $1/\lambda  $. Such potentials can be
viewed as elements of a Kac-Moody algebra $ \widehat{\fr{g}}_C $.
Roughly speaking the construction of $\widehat{\fr{g}}_C $ involves a
simple Lie algebra $\fr{g} $ and an automorphism $C $ of finite order,
i.e. there exist such an integer $h $ that $C^h =\openone  $. Then we can
split $\fr{g} $ into a direct sum of linear subspaces
\begin{equation}\label{eq:5.1}
\fr{g} = \opl_{k=0}^{h-1} \fr{g}^{(k)},
\end{equation}
which are eigensubspaces of $C $, i.e. if
\begin{equation}\label{eq:5.1'}
X^{(k)} \in \fr{g}^{(k)} \qquad \Leftrightarrow \qquad
C(X^{(k)}) =\omega ^{-k} X^{(k)},
\end{equation}
where $\omega =\exp(2\pi i/h) $. The decomposition (\ref{eq:5.1})
satisfies the grading condition:
\begin{equation}\label{eq:5.2}
\left[ X^{(k)}, X^{(m)}\right] = X^{(k+m)} \in \fr{g}^{(k+m)}.
\end{equation}
where the superscript $k+m $ in $\fr{g}^{(k+m)} $ is understood modulo $h
$.  Then the elements of the corresponding Kac-Moody algebra $
\widehat{\fr{g}}_C  $ have the form:
\begin{equation}\label{eq:6.1}
X(\lambda ) = \sum_{k\leq N_1}^{} \lambda ^k X^{(k)}, \qquad X^{(k)} \in
\fr{g}^{(k)},
\end{equation}
Obviously due to (\ref{eq:5.2}) the commutator of any two elements
$X(\lambda ) $, $Y(\lambda ) $ of the form (\ref{eq:6.1}) will also have
the form (\ref{eq:6.1}).

The classification and the theory of the Kac-Moody algebras can be
found in \cite{Kac,Helg}. Their simplest realization can be
obtained from a pair  ($\fr{g}, C $) with a few special choices of
the automorphism of finite order $C $, namely:

a) $C=\openone  $; then each of the subspaces $\fr{g}^{(k)}\simeq \fr{g}
$. This leads to a generic GZS system if $J $ is real and to a generic CBC
system if $J $ is complex.

b) $C^h=1 $ where $C $ is the Coxeter automorphism of $\fr{g} $ and
$h$ is the Coxeter number. This leads to a CBC system with $\bbbz_h
$-reduction and will be used in analyzing the NLEE (\ref{eq:3.2}) and
(\ref{eq:3.3}).

c) $CV $ where $V $ is a nontrivial external automorphism of $\fr{g}$.
Such gradings also lead to interesting NLEE but will not be used in this
paper.

The Kac-Moody algebras are obtained from the constructions a)--c) with
additional central extensions; they are split into three classes: of
height 1 (cases a) and b)) and of height 2 and 3  depending on the order of
$V $.

The potential $U(x,t,\lambda ) $ for the $N $-wave equations equals
$[J,Q(x,t)]-\lambda J $ belongs to a Kac-Moody algebra with $\fr{g}\simeq
sl(n) $ and $C=\openone $. The potential $\tilde{U}(x,t,\lambda
)=\tilde{q}(x,t)-\lambda \tilde{J} $ of the form (\ref{eq:4.2}) and
(\ref{eq:To3}) gives rise to the NLEE (\ref{eq:3.2}) and (\ref{eq:3.3})
is related to Kac-Moody algebra of the class b) with $\fr{g}\simeq sl(n)
$. The Coxeter number then is $h=n $; the Coxeter automorphism can be
realized as inner automorphism of the form:

\begin{equation}\label{eq:7.1}
C(X) = C_0 X C_0^{-1}, \qquad C_0=\sum_{k=1}^{n} \omega^k E_{kk}, \qquad
\omega =e^{2\pi i/n},
\end{equation}
where $C $  obviously satisfies $C^n=\openone  $. With this choice
of $C $ we can easily check that the linear subspaces
$\fr{g}^{(k)} $ are spanned by
\begin{equation}\label{eq:7.2}
\fr{g}^{(k)} \equiv \mbox{l.c.\;} \left\{ E_{j,j+k}, \quad j,k=1,\dots , n
\right\},
\end{equation}
and $j+k $ is considered modulo $n $. Comparing (\ref{eq:4.2}),
(\ref{eq:To2}) with (\ref{eq:7.2}) below we find that $\tilde{q}(x,t)\in
\fr{g}^{(0)} $ and $\tilde{J}\in \fr{g}^{(1)} $.  Note that now the
condition $X^{(k)}\in \fr{g}^{(k)} $ imposes a set of nontrivial
constraints on $X^{(k)} $.

The idea to use finite order automorphisms for the reductions of the NLEE
was proposed first by Mikhailov \cite{Mikh} who introduced also the notion
of the reduction group. The $\bbbz_n $-reduction condition according to
\cite{Mikh} is introduced by:
\begin{equation}\label{eq:7.3}
C (\tilde{U}(x,t,\lambda \omega )) = \tilde{U}(x,t,\lambda ),\qquad
C (\tilde{V}(x,t,\lambda \omega )) = \tilde{V}(x,t,\lambda ),
\end{equation}
where we have chosen the simplest possible realization of the
$\bbbz_n $ group on the complex $\lambda  $-plane: $\lambda \to
\lambda \omega  $ with $\omega =\exp(2\pi i/n) $.

The Kac-Moody algebras, like the semi-simple Lie algebras have an
important property which ensures the solvability of the inverse
scattering problem for $L(\lambda ) $ and the non-degeneracy of
the Hamiltonian structures of the NLEE. While the semi-simple Lie
algebras possess just one invariant bilinear form
$\boldsymbol{\langle} X,Y \boldsymbol{ \rangle} \equiv \tr \left(
\ad_X, \ad_Y \right) $ the Kac-Moody algebras possess a family
of invariant bilinear forms:
\begin{equation}\label{eq:Kill}
\bigla X(\lambda ),Y(\lambda ) \bigra^{(p)}= \res_{\lambda =0}
\lambda ^{-p-1} \boldsymbol{\langle} X(\lambda ),Y(\lambda )
\boldsymbol{\rangle} ,
\end{equation}
for all integer values of $p $.

We will need also a central extension of the Kac-Moody algebras
$\widetilde{\fr{g}}=\widehat{\fr{g}}\oplus c $ where the central element
is generated by the 2-cocycle:
\begin{equation}\label{eq:c-ex}
\boldsymbol{\omega }_p (X(x,\lambda ),Y(x,\lambda )) = \int_{-\infty}
^{\infty } dx \Bigla X(x,\lambda ), {d Y(x,\lambda ) \over dx} \Bigra^{(p)}.
\end{equation}
This means that each element of $\tilde{\fr{g}} $ is a pair $(X(x,\lambda
),c_X) $ where $c_X $ is a constant. The commutation relation in
$\tilde{\fr{g}} $ is defined by:
\begin{equation}\label{eq:g-c}
\left[ (X(x,\lambda ), c_X), (Y(x,\lambda ), c_Y) \right] =
\left( [X(x,\lambda ), Y(x,\lambda )],  \boldsymbol{\omega }_p
(X(x,\lambda ),Y(x,\lambda ))\right) .
\end{equation}
Important role for the Hamiltonian formulation of the NLEE is played by
the dual algebras $\hat{\fr{g}}^* $, $\tilde{\fr{g}}^*
=\hat{\fr{g}}^*\oplus c $ and their splittings into direct sums of
Borel-like subalgebras. These splittings for $\hat{\fr{g}} =
\hat{\fr{g}}_+ \oplus \hat{\fr{g}}_- $ look like:
\begin{eqnarray}\label{eq:borel}
\hat{\fr{g}}_+ \equiv \left\{ \sum_{k=0}^{N_1} u_k(x)\lambda ^k \right\},
\qquad  \hat{\fr{g}}_- \equiv \left\{ \sum_{k=-\infty }^{-1}
u_k(x)\lambda ^k \right\},
\end{eqnarray}
and for the dual $\hat{\fr{g}}^* = \hat{\fr{g}}_+^* \oplus
\hat{\fr{g}}_-^* $:
\begin{eqnarray}\label{eq:borel*}
\hat{\fr{g}}_+^* \equiv \left\{ \sum_{k=-N_1 }^{p} u_k(x)\lambda ^k
\right\}, \qquad  \hat{\fr{g}}_-^* \equiv \left\{ \sum_{k=p+1}^{\infty }
u_k(x)\lambda ^k \right\}.
\end{eqnarray}
The co-adjoint orbits of $\tilde{\fr{g}} $  on $\tilde{\fr{g}}^* $ in fact
are isomorphic to the space of coefficients for which the NLEE is written.
Thus they are natural candidate for the phase space of these equations.
The freedom provided by the parameter $p $ is directly related
to the existence of hierarchy of Hamiltonian structures for the NLEE.

\section*{Fundamental analytic solutions}\label{sec:In.3}

The second important tool in this scheme is the fundamental analytic
solution (FAS) of $L(\lambda ) $. We will see that using the FAS one is
able to:

-- reduce the solving of the ISP for $L(\lambda ) $ to an equivalent
Riemann-Hilbert problem (RHP) for the FAS \cite{Sha,ZMNP,ZaSha};

-- construct the kernel of the resolvent for $L(\lambda ) $ and derive the
spectral decomposition for $L(\lambda ) $ \cite{GeKu1,Ge2,GeYa};

-- construct the `squared' solutions of $L(\lambda ) $ which
allow the interpretation of the ISM as a generalized Fourier transform
(GFT) \cite{AKNS,Kau*76,KauNew,GeKh,GeYa*85,GeYa*86,GeYa};

-- construct the Green function for the recursion operators
$\Lambda _\pm $ and prove the completeness relation for the
`squared' solutions. This property ensures the uniqueness of the
solution of the ISM \cite{GeKh,GeKu,Ge,GeYa}.

The existence of FAS is ensured by the
analytic dependence of both $U(x,t,\lambda ) $ and $V(x,t,\lambda ) $ on $
\lambda  $. The properties of FAS depend crucially on the boundary
conditions imposed on the potential $q(x,t) $. For simplicity here we
consider the class of potentials $q(x,t) $ that are sufficiently smooth
functions of $x $ and tend to zero fast enough for $x\to\pm\infty  $ for
any fixed value of $t $.

The FAS for the  Zakharov-Shabat system (i.e. $\fr{g}\simeq sl(2) $) can
easily be constructed due to the fact that each of the columns of the Jost
solutions
\begin{equation}\label{eq:Jost}
L(\lambda )\psi _\pm (x,t,\lambda ) =0, \qquad
\lim_{x\to\pm\infty }  e^{iJ \lambda x} \psi _\pm (x,t,\lambda ) =
\openone  ,
\end{equation}
allow analytic extension either for $\lambda \in \bbbc_+ $ or for $\lambda
\in \bbbc_- $, see \cite{AKNS}.

If we analyze the analyticity properties of the Jost solutions
$\psi _\pm(x,t,\lambda ) $ related to algebras of higher rank one finds
that only the first and the last columns of $\psi _\pm(x,t,\lambda ) $
allow analytic extensions off the real $\lambda $-axis. An important
result of Zakharov and Manakov \cite{ZaMa,ZMNP} consisted in showing that
a FAS for the GZS with $\fr{g}\simeq sl(n) $ and real-valued $J $ can be
constructed by taking proper linear combinations of the columns of $\psi
_\pm(x,t,\lambda ) $.

The construction is more complicated for the Caudrey-Beals-Coifman (CBC)
systems when the eigenvalues of $J $ are complex  \cite{BeCo1,BeCo2,Caud}.
The generalization of this construction  for CBC systems related to any
simple Lie algebra $\fr{g} $ was done in \cite{GeYa}.

We make attempt to outline the construction and the properties
of each of these tools. Then we show how the FAS can be used to construct
the kernel of the resolvent of $L(\lambda ) $  and to exhibit its
spectral properties and the structure of its discrete spectrum.
Finally we illustrate how these tools can be used in the analysis of
the NLEE and their fundamental properties and finish with some
conclusions.

Both these aspects are rather broad; they have been widely discussed in
hundreds of papers. Therefore inevitably the list of references consists
mainly of reviews and monographs and bears an illustrative character. The
thorough reader is advised to consult also the papers referred to in these
references.

\specialsection{Construction of the FAS}\label{sec:2}

\section*{Preliminaries: Jost solutions and scattering
matrix}\label{sec:2.1}

The direct and the inverse scattering problem for the Lax operator
(\ref{eq:2.2}) will be done for fixed $t $ and in most of the
corresponding formulae $t $ will be omitted.

The crucial fact that determines the spectral properties of the operator
$L $ is the choice of the class of functions where from we shall choose
the potential $q(x) $. Below for simplicity we assume that the
potential $q(x) $ is defined on the whole axis and satisfies the following
conditions:

\begin{description}

\item [C.1] By $q(x)\in {\cal  M}_S $ we mean that $q(x) $
possesses smooth derivatives of all orders and  falls off to zero for
$|x|\to\infty $ faster than any power of $x $:
\[
\lim_{x\to\pm\infty } |x|^k q(x) =0, \qquad \forall k=0,1,2,\dots
\]

\item [C.2] $q(x) $ is such that  the corresponding operator $L $ has only
a finite  number of simple discrete eigenvalues.

\end{description}

Below we will use along with $L\psi (x,\lambda )=0 $
also the following equivalent formulations of the system (\ref{eq:2.2}):
\begin{eqnarray}\label{eq:N16.1}
i {d\xi  \over dx } + q(x,t) \xi (x,\lambda ) - \lambda [J,\xi
(x,\lambda )] = 0, &\quad &\xi(x,\lambda ) = \psi(x,\lambda )e^{i\lambda
Jx}, \\
\label{eq:N16.2}
i {d\hat{\psi } \over dx } - \hat{\psi } (x,\lambda ) q(x,t) + \lambda
\hat{\psi} (x,\lambda )J = 0,& \quad &\hat{\psi }(x,\lambda ) =
(\psi(x,\lambda))^{-1}, \\
\label{eq:N16.3}
i {d\hat{\xi}  \over dx } -\hat{\xi} (x,\lambda ) q(x,t) + \lambda
[\hat{\xi} (x,\lambda ),J] = 0,& \quad &\hat{\xi}(x,\lambda ) =
e^{-i\lambda Jx} \hat{\psi}(x,\lambda ).
\end{eqnarray}
where by `hat' we denote the inverse matrix, $\hat{X}\equiv X^{-1} $.
The Jost solutions $\xi_\pm(x,\lambda ) $, $\hat{\chi }_\pm(x,\lambda ) $
and $\hat{\xi}_\pm(x,\lambda ) $ for the systems
(\ref{eq:N16.1})--(\ref{eq:N16.3}) can be introduced by:
\begin{eqnarray*}
&& \lim_{x\to\pm\infty }\xi_\pm(x,\lambda )=\openone , \qquad
\lim_{x\to\pm\infty }\psi_\pm(x,\lambda )e^{-i\lambda Jx} =\openone ,
\qquad  \lim_{x\to\pm\infty }\hat{\xi}_\pm(x,\lambda )=\openone ,
\end{eqnarray*}
in analogy to (\ref{eq:10.2}); then their scattering matrices are:
\begin{eqnarray*}
&& T_2(\lambda ) \equiv e^{i\lambda Jx}(\xi_-(x,\lambda ))^{-1}
\xi_+(x,\lambda )e^{-i\lambda Jx} =T(\lambda ), \\
&& T_3(\lambda ) \equiv \hat{\psi}_+(x,\lambda ) (\hat{\psi}_-(x,\lambda
))^{-1} = T^{-1}(\lambda ), \\
&& T_4(\lambda ) \equiv e^{i\lambda Jx} \hat{\xi}_+(x,\lambda )
(\hat{\xi}_-(x,\lambda ))^{-1}e^{-i\lambda Jx}= T^{-1}(\lambda ),
\end{eqnarray*}

Below we will consider two specific reductions of the Lax operator: the
GZSs with $\bbbz_2 $-reduction:
\begin{equation}\label{eq:Z_2}
B(U^{\dagger}(x,t,\epsilon \lambda^*))B^{-1}= U(x,t,\lambda ), \qquad B^2
=\openone, \qquad \epsilon =\pm 1.
\end{equation}
The first possible choice for $B=\diag (\epsilon _1,\dots,\epsilon _n) $,
$\epsilon _j=\pm 1 $ with $\epsilon =1 $ leads to the classical $N $-wave
equations \cite{ZaMa,ZMNP} with
\begin{equation}\label{eq:N-w-Z2}
q_{kj}^*(x,t)=\epsilon _k\epsilon _jq_{jk}(x,t), \qquad J=
\diag(a_1,\dots, a_n), \qquad a_k= \epsilon a_k^*.
\end{equation}
Since all eigenvalues of $J $ are real ($\epsilon =1) $, or purely
imaginary ($\epsilon =-1) $, the Lax operator becomes a GZSs.  The second
choice for $B $:
\begin{equation}\label{eq:Z-N}
B=\sum_{k=1}^{n} E_{k\bar{k}}, \qquad \bar{k}=n+1-k, \qquad
\epsilon =-1,
\end{equation}
will be used in combination with the $\bbbz_n $-reduction:
\begin{equation}\label{eq:Z_n}
\tilde{C}_0(U(x,t,\omega \lambda ))\tilde{C}_0^{-1}= U(x,t,\lambda ),
\qquad \tilde{C}^n =\openone .
\end{equation}
which leads to the CBC system. For the sake of convenience in
doing the spectral problem of CBCs we choose  $C_0=\sum_{k=1}^{n}
E_{k,k+1} $; then $L(\lambda ) $ has the form (\ref{eq:2.2}) with
diagonal complex-valued $J $ given by (\ref{eq:4.2}) or
(\ref{eq:To3}) where $c_0=1 $ (resp. $c_0=i $) if $\epsilon =1 $
(resp. $\epsilon =-1 $). Both Lax operators will have similar spectral
properties.

In solving the NLEE (\ref{eq:3.2}) and (\ref{eq:3.3}) we will need to
apply both reductions (\ref{eq:Z_2}) and (\ref{eq:Z_n}) simultaneously.
An attempt for classification of the $\bbbz_2 $-reductions is made
in \cite{GGKI}.

\section*{The FAS of the GZSs with $\bbbz_2 $-reduction.}\label{sec:2.2}

Let us outline without proofs the construction of the FAS for the
GZSs with real $J $, see \cite{ZaMa,ZMNP,BeCo1,Caud,GeYa}. For
definiteness we assume that the real eigenvalues of $J $ are
pair-wise different and ordered as follows:
\begin{equation}\label{eq:J-real}
J=\diag(a_1,\dots, a_n), \qquad a_1>a_2 >\dots >a_n.
\end{equation}

\begin{proposition}\label{pro:2}
Let the potential of (\ref{eq:2.2}) $q(x)\in{\cal  M}_S $ satisfies
conditions (C.1), (C.2) and (\ref{eq:N-w-Z2}).  Then:

a) the Jost solutions $\xi_ \pm (x,\lambda ) $ and $\hat{\xi}_ \pm
(x,\lambda ) $ of (\ref{eq:N16.1}),  (\ref{eq:N16.2}) exist and are well
defined functions for $\lambda \in \bbbr $;

b) the matrix elements of the scattering matrix $T(\lambda ) $ and its
inverse $\hat{T}(\lambda ) $ are Schwartz-type functions of $\lambda  $
for $\lambda \in\bbbr $.

\end{proposition}

\begin{remark}\label{rem:1}
The proposition \ref{pro:2} concerns the Jost solutions as fundamental
solutions. One can prove that the first and the last columns
$\xi_\pm^{[1]}(x,\lambda ) $ and $\xi_\pm^{[n]}(x,\lambda )$  of the Jost
solutions allow analytic extension with respect to $\lambda $ as follows:
\[
\arraycolsep=6pt
\begin{array}{lcccc}
\text{Column } & \xi_+^{[1]}(x,\lambda ) & \xi_+^{[n]}(x,\lambda ) &
\xi_-^{[1]}(x,\lambda ) & \xi_-^{[n]}(x,\lambda ) \\
\text{Analytic for } & \lambda \in \bbbc_- & \lambda \in \bbbc_+
& \lambda \in \bbbc_+ & \lambda \in \bbbc_- \end{array},
\]
Analogously the first and the last rows of the Jost solutions
$\hat{\xi}_\pm^{[1]}(x,\lambda ) $ and  $\hat{\xi}_\pm^{[n]}(x,\lambda )$
allow analytic extension with respect to $\lambda  $ as follows:
\[
\arraycolsep=6pt
\begin{array}{lcccc}
\text{Row } & \hat{\xi}_+^{[1]}(x,\lambda ) &
\hat{\xi}_+^{[n]}(x,\lambda ) & \hat{\xi}_-^{[1]}(x,\lambda ) &
\hat{\xi}_-^{[n]}(x,\lambda ) \\
\text{Analytic for } & \lambda \in \bbbc_+ &
\lambda \in \bbbc_- & \lambda \in \bbbc_- & \lambda \in \bbbc_+
\end{array},
\]
All the other columns of $\xi_\pm(x,\lambda ) $ and rows of
$\hat{\xi}_\pm(x,\lambda ) $ are defined only for $\lambda \in\bbbr $
and do not allow analytic extensions off the real axis.

\end{remark}

We start by explaining the construction of the FAS $\chi ^\pm(x,\lambda )
$ or rather of the solutions
\begin{equation}\label{eq:s6.1}
\xi^\pm(x,\lambda )=\chi^\pm(x,\lambda )e^{i\lambda Jx} .
\end{equation}
to equation (\ref{eq:N16.1}) which allow analytic extensions for $\lambda
\in\bbbc_\pm $. Skipping the details (see \cite{Sha,ZMNP,ZaMa}) we
formulate the answer and determine $\xi^+(x,\lambda ) $ as the solution of
the following set of integral equations:
\begin{subequations}\label{eq:12.1}
\begin{eqnarray}\label{eq:12.1a}
&&\xi^+_{ij}(x,\lambda )= \delta _{ij} + i \int_{-\infty }^{x} dy
e^{-i\lambda (a_i-a_j)(x-y)} \sum_{p=1}^{h} q_{ip}(y)
\xi^+_{pj}(y,\lambda ), \qquad i\geq j ;\\
\label{eq:12.1b}
&&\xi^+_{ij}(x,\lambda )=  i \int_{\infty }^{x} dy e^{-i\lambda
(a_i-a_j)(x-y)} \sum_{p=1}^{h} q_{ip}(y) \xi^+_{pj}(y,\lambda ), \qquad
i<j ;
\end{eqnarray}
\end{subequations}
Analogously we define $\xi^-(x,\lambda ) $ as the solution of the set of
integral equations:
\begin{subequations}\label{eq:12.2}
\begin{eqnarray}\label{eq:12.2a}
&&\xi^-_{ij}(x,\lambda )= i \int_{\infty }^{x} dy
e^{-i\lambda (a_i-a_j)(x-y)} \sum_{p=1}^{h} q_{ip}(y)
\xi^-_{pj}(y,\lambda ), \qquad i>j ;\\
\label{eq:12.2b}
&&\xi^-_{ij}(x,\lambda )= \delta _{ij} +  i \int_{-\infty }^{x} dy
e^{-i\lambda (a_i-a_j)(x-y)} \sum_{p=1}^{h} q_{ip}(y)
\xi^-_{pj}(y,\lambda ), \qquad i\leq j ;
\end{eqnarray}
\end{subequations}

\begin{theorem}\label{th:6.1}
Let $q(x)\in {\cal  M}_S $ satisfies conditions (C.1), (C.2) and let $J $
satisfy (\ref{eq:J-real}).  Then the solution $\xi^+(x,\lambda ) $ of the
eqs. (\ref{eq:12.1}) (resp.  $\xi^-(x,\lambda ) $ of the eqs.
(\ref{eq:12.2})) exists and allows analytic extension for $\lambda \in
\bbbc_+ $ (resp. for $\lambda \in \bbbc_- $).
\end{theorem}

\begin{remark}\label{rem:6.1}
Due to the fact that in eq. (\ref{eq:12.1}) we have both $\infty  $ and
$-\infty  $ as lower limits the equations are rather of Fredholm than of
Volterra type. Therefore we have to consider also the Fredholm
alternative, i.e. there may exist finite number of values of
$\lambda=\lambda _k^\pm\in\bbbc_\pm $ for which the solutions
$\xi^\pm(x,\lambda ) $ have zeroes and pole singularities in $\lambda $.
The points $\lambda _k^\pm $ in fact are the discrete eigenvalues of
$L(\lambda ) $ in $\bbbc_\pm $.  \end{remark}

The reduction condition (\ref{eq:Z_2}) with $\epsilon =1 $ means that the
FAS and the scattering matrix $T(\lambda ) $ satisfy:
\begin{eqnarray}\label{eq:N24.0}
B\left(\chi ^+(x,\lambda ^*)\right)^\dag B^{-1}= (\chi ^-(x,\lambda
))^{-1}, \qquad
B\left( T(\lambda^*)\right)^\dag B^{-1} = (T(\lambda))^{-1}.
\end{eqnarray}

Each fundamental solution of the Lax operator is uniquely determined by
its asymptotic for $x\to\infty  $ or $x\to -\infty  $.  Therefore in order
to determine the linear relations between the FAS and the Jost solutions
for $\lambda \in \bbbr $ we need to calculate the asymptotics of FAS for
$x\to\pm\infty $.  Taking the limits in the right hand sides of the
integral equations (\ref{eq:12.1}) and (\ref{eq:12.2}) we get:
\begin{subequations}\label{eq:14.1}
\begin{eqnarray}\label{eq:14.1a}
\qquad \lim_{x\to -\infty } \xi _{ij}^{+}(x,\lambda ) = \delta _{ij},
\quad \text{for}\; i\geq j; \qquad \lim_{x\to \infty } \xi
_{ij}^{+}(x,\lambda ) = 0, \quad \text{for}\; i< j;\\
\label{eq:14.1b}
\qquad \lim_{x\to -\infty } \xi _{ij}^{-}(x,\lambda ) = \delta _{ij},
\quad \text{for}\; i\leq j; \qquad \lim_{x\to \infty } \xi
_{ij}^{-}(x,\lambda ) = 0, \quad \text{for}\; i> j;
\end{eqnarray}
\end{subequations}
This can be written in compact form using (\ref{eq:s6.1}):
\begin{equation}\label{eq:14.3}
\chi ^\pm(x,\lambda ) = \psi _-(x,\lambda ) S^\pm (\lambda )
= \psi _+(x,\lambda ) T^\mp (\lambda ) D^\pm (\lambda ) ,
\end{equation}
where the matrices $S^\pm (\lambda )  $, $D^\pm (\lambda )  $ and  $T^\pm
(\lambda )  $ are of the form:
\begin{subequations}\label{eq:14.4}
\begin{eqnarray}\label{eq:14.4a}
&&\qquad  S^+(\lambda ) = \left( \begin{array}{cccc}
1 & S_{12}^{+} & \dots & S_{1n}^{+} \\
0 & 1 & \dots & S_{2n}^{+} \\ \vdots & \vdots & \ddots & \vdots \\
0 & 0 & \dots & 1 \end{array} \right),
\qquad T^+(\lambda ) = \left( \begin{array}{cccc}
1 & T_{12}^{+} & \dots & T_{1n}^{+} \\
0 & 1 & \dots & T_{2n}^{+} \\ \vdots & \vdots & \ddots & \vdots \\
0 & 0 & \dots & 1 \end{array} \right), \\
\label{eq:14.4b}
&&\qquad  D^+(\lambda ) = \diag (D_1^+, D_2^+, \dots, D_n^+), \qquad
D^-(\lambda ) = \diag (D_1^-, D_2^-, \dots, D_n^-), \\
\label{eq:14.4c}
&&\qquad  S^-(\lambda ) = \left( \begin{array}{cccc}
1 & 0 & \dots & 0 \\
S_{21}^{-} & 1 & \dots & 0 \\ \vdots & \vdots & \ddots & \vdots \\
S_{n 1}^{-} & S_{n 2}^{-}
& \dots & 1 \end{array} \right), \qquad
T^-(\lambda ) = \left( \begin{array}{cccc}
1 & 0 & \dots & 0 \\
T_{21}^{-} & 1 & \dots & 0 \\ \vdots & \vdots & \ddots & \vdots \\
T_{n 1}^{-} & T_{n 2}^{-} & \dots & 1 \end{array} \right),
\end{eqnarray}
\end{subequations}

Let us now relate the factors $T^\pm(\lambda ) $, $S^\pm(\lambda ) $ and
$D^\pm(\lambda ) $ to the scattering matrix $T(\lambda ) $. Comparing
(\ref{eq:14.3}) with (\ref{eq:10.3}) we find
\begin{equation}\label{eq:15.1}
T(\lambda ) = T^-(\lambda )D^+(\lambda ) \hat{S}^+(\lambda )
= T^+(\lambda )D^-(\lambda ) \hat{S}^-(\lambda ) ,
\end{equation}
i.e. $T^\pm(\lambda ) $, $S^\pm(\lambda ) $ and $D^\pm(\lambda ) $ are the
factors in the Gauss decomposition of $T(\lambda ) $.

It is well known how given $T(\lambda ) $ one can construct explicitly its
Gauss decomposition, see the Appendix \ref{sec:A1}. Here we need
only the expressions for $D^\pm(\lambda ) $:
\begin{equation}\label{eq:15.2}
D^+_{j}(\lambda ) = {m_j^+(\lambda )  \over m^+_{j-1}(\lambda ) }, \qquad
D^-_{j}(\lambda ) = {m_{n-j+1}^-(\lambda )  \over m^-_{n-j}(\lambda )},
\end{equation}
where $m_j^\pm $ are the principal upper and lower minors of $T(\lambda )
$ of order $j $.

\begin{corollary}\label{cor:7.1}
The upper (resp. lower) principal minors $m_j^\pm(\lambda ) $ (resp.
$m_{j}^-(\lambda ) $ of $T(\lambda ) $ are analytic functions of $\lambda$
for $\lambda \in\bbbc_+ $ (resp.  for $\lambda \in\bbbc_- $).

\end{corollary}

\begin{proof}
{}Follows directly from theorem \ref{th:6.1}, from the limits:
\begin{equation}\label{eq:lim-Dj}
\lim_{x\to\infty } \xi^+_{jj}(x,\lambda ) = D_j^+(\lambda ), \qquad
\lim_{x\to \infty } \xi^-_{jj}(x,\lambda ) = D_j^-(\lambda ),
\end{equation}
and from (\ref{eq:14.4b}) and (\ref{eq:15.2}).
\end{proof}

\begin{corollary}\label{cor:7.2} The following relations hold:
\[
\mbox{a)} \lim_{\lambda \to\infty } \xi ^\pm (x,\lambda ) =\openone  ;
\qquad \mbox{b)} \lim_{\lambda \to\infty } m_j^\pm (\lambda ) =1 .
\]
\end{corollary}

\begin{proof}{}

a) follows from the integral equations (\ref{eq:12.1}), (\ref{eq:12.2})
taking into account that the integrands in their right hand sides vanish
for $\lambda \to\infty  $. b) follows from a), (\ref{eq:lim-Dj}) and
(\ref{eq:14.4b}).
\end{proof}

In what follows we will also assume that the set of minors
$m_k^\pm(\lambda ) $ have only finite number of simple zeroes located at
the points
\begin{equation}\label{eq:Zero}
{\cal  Z} \equiv \left\{ \lambda _j^\pm \in \bbbc_\pm , \quad j=1,\dots,
N.\right\}
\end{equation}
Generically each of the $\lambda _j^\pm $ can be a zero of a string of
minors, e.g.:
\begin{equation}\label{eq:N25.2}
m_k^+(\lambda ) = (\lambda -\lambda _j^+)\dot{m}_{k,j}^+ + {\cal  O}\left(
(\lambda -\lambda _j^+)^2\right),
\end{equation}
for $1\leq I_j  <F_j \leq n $. Let us introduce the quantities
$b_{jk} $ as follows:
\begin{equation}\label{eq:N25.6}
b_{jk}= \left\{ \begin{array}{ll} 1 \qquad & \text{if $\lambda _j^+ $ is a
zero of $m_k^+(\lambda ) $;} \\ 0 \qquad & \text{if $\lambda _j^+ $ is not
a zero of $m_k^+(\lambda ) $.} \end{array} \right.
\end{equation}
and note that the reduction (\ref{eq:Z_2}) means that the
Gauss factors of $T(\lambda ) $ satisfy ($\epsilon =1$):
\begin{subequations}\label{eq:N24.5}
\begin{eqnarray}\label{eq:N24.5a}
&& B\left( \hat{S}^+(\lambda ^*)\right)^\dag B^{-1} = S^-(\lambda ), \qquad
B\left( \hat{T}^+(\lambda ^*)\right)^\dag B^{-1} = T^-(\lambda ), \\
\label{eq:N24.5b}
&& \left( \hat{D}^+(\lambda ^*)\right)^\dag = D^-(\lambda ).
\end{eqnarray}
\end{subequations}
The relations (\ref{eq:N24.5a}) are strictly valid only for $\lambda \in
\bbbr $ while (\ref{eq:N24.5b}) together with (\ref{eq:14.4b}) and
(\ref{eq:15.2}) leads to the following constraints on the minors
$m_k^\pm(\lambda ) $:
\begin{equation}\label{eq:N24.6}
\left( m_k^+(\lambda ^*)\right)^* = m_{n-k}^-(\lambda ).
\end{equation}
Thus if $\lambda _k^+ $ is a zero of $m_k^+(\lambda ) $ then
$\lambda _k^-=(\lambda _k^+)^* $ is a zero of $m_{n-k}^-(\lambda ) $.

\section*{The FAS of the CBCs with $\bbbz_n $-reduction.}\label{sec:2.3}

The crucial difference with the $\bbbz_2 $-case treated above
consists in the fact that now $J $ is given by (\ref{eq:4.2}) or
(\ref{eq:To3}) and has complex eigenvalues.  Skipping the details
(see \cite{BeCo1,BeCo2,Caud,GeYa}) we just outline the procedure
of constructing the FAS.

{}First we have to determine the regions of analyticity. For
potentials $q(x) $ satisfying the conditions (C.1) and (C.2) and
subject to the $\bbbz_n $-reduction (\ref{eq:Z_n}) these regions
are the $2n $ sectors $ \Omega _\nu  $ separated by the rays
$l_\nu  $ on which $\im \lambda (a_j-a_k) =0 $. We remind that if
we assume also the $\bbbz_2 $-reduction (\ref{eq:Z-N}) with
$c_0^*=\epsilon c_0 $ then $a_k=c_0\omega ^{k-1/2} $. Then the
rays $l_\nu $ are given by:
\begin{eqnarray}\label{eq:l-nu}
l_\nu \colon \arg(\lambda) =\phi _0 + {\pi (\nu -1) \over n } ,
\qquad \nu =1,\dots , 2n,
\end{eqnarray}
where $\phi _0=\pi/(2n) $ only if $\epsilon =1 $ and $n $ is odd;
in all other cases $\phi _0=0 $. Thus the neighboring rays $l_\nu
$ and $l_{\nu +1} $ close angles equal to $\pi/n $.

The next step is to construct the set of integral equations analogous to
(\ref{eq:12.1}) whose solution will be analytic in $\Omega _\nu  $. To
this end we associate with each sector $\Omega _\nu  $ the relations
(orderings) $\gnu $ and $\lnu $ by:
\begin{equation}\label{eq:gnu}
\begin{array}{c} i\gnu j \\ i\lnu j \end{array} \qquad \mbox{if} \qquad
\begin{array}{ll} \im \lambda (a_i-a_j) <0 & \quad \mbox{for } \lambda \in
\Omega _\nu , \\ \im \lambda (a_i-a_j) >0 & \quad \mbox{for } \lambda \in
\Omega _\nu  . \end{array}
\end{equation}
Then the solution of the system (\ref{eq:12.1})
\begin{subequations}\label{eq:12.1nu}
\begin{eqnarray}\label{eq:12.1nua}
&&\xi^\nu _{ij}(x,\lambda )= \delta _{ij} + i \int_{-\infty }^{x} dy
e^{-i\lambda (a_i-a_j)(x-y)} \sum_{p=1}^{h} q_{ip}(y)
\xi^\nu _{pj}(y,\lambda ), \qquad i\genu j ;\\
\label{eq:12.1nub}
&&\xi^\nu _{ij}(x,\lambda )=  i \int_{\infty }^{x} dy e^{-i\lambda
(a_i-a_j)(x-y)} \sum_{p=1}^{h} q_{ip}(y) \xi^\nu _{pj}(y,\lambda ), \qquad
i\lnu j ;
\end{eqnarray}
\end{subequations}
will be the FAS of the CBCs in the sector $\Omega _\nu  $. The asymptotics
of $\xi^\nu (x,\lambda ) $ and $\xi^{\nu -1} (x,\lambda )  $ along the
ray $l_\nu  $ can be written in the form:
\begin{subequations}\label{eq:xi-as}
\begin{eqnarray}\label{eq:xi-as.a}
\lim_{x\to -\infty } e^{i\lambda Jx} \xi^\nu (x,\lambda e^{i0} )
e^{-i\lambda Jx} &=& S_\nu ^+ (\lambda ), \qquad \lambda \in l_\nu ,\\
\label{eq:xi-as.b}
\lim_{x\to -\infty } e^{i\lambda Jx} \xi^{\nu -1} (x,\lambda e^{-i0} )
e^{-i\lambda Jx} &=& S_{\nu} ^- (\lambda ), \qquad \lambda \in l_{\nu },\\
\label{eq:xi-as.c}
\lim_{x\to \infty } e^{i\lambda Jx} \xi^\nu (x,\lambda e^{i0} )
e^{-i\lambda Jx} &=& T_\nu ^-D_\nu ^+ (\lambda ), \qquad \lambda \in l_\nu
,\\
\label{eq:xi-as.d}
\lim_{x\to \infty } e^{i\lambda Jx} \xi^{\nu -1} (x,\lambda e^{-i0} )
e^{-i\lambda Jx} &=& T_{\nu} ^+D_{\nu } ^-(\lambda ), \qquad \lambda \in
l_{\nu } ,
\end{eqnarray}
\end{subequations}
where the matrices $S_\nu ^+ $, $T_\nu ^+ $ (resp. $S_\nu ^- $, $T_\nu ^-
$) are upper-triangular (resp. lower-triangular) with respect to the $\nu
$-ordering. As in the previous case they provide the Gauss decomposition
of the scattering matrix with respect to the $\nu  $-ordering, i.e.:
\begin{equation}\label{eq:nu-gauss}
T_\nu (\lambda ) = T_\nu ^-(\lambda ) D_\nu ^+(\lambda ) \hat{S}_\nu ^+
(\lambda ) =  T_\nu ^+(\lambda ) D_\nu ^-(\lambda ) \hat{S}_\nu ^-
(\lambda ) , \qquad  \lambda \in l_\nu .
\end{equation}
More careful analysis shows \cite{GeYa} that in fact $T_\nu
(\lambda ) $ belongs to a subgroup ${\cal G}_\nu$ of $SL(n,\bbbc)
$. Indeed, with each ray $l_\nu $ one can relate a subalgebra
$\fr{g}_\nu \subset sl(n,\bbbc) $.

If $\bbbz_n $-symmetry is present each of these subalgebras
$\fr{g}_\nu  $ is a direct sum of $sl(2) $-subalgebras. Each such
$sl(2) $-subalgebra can be specified by a pair of indices $(k,s) $
and is generated by:
\begin{equation}\label{eq:sl2-gen}
h^{(k,s)} =E_{kk}-E_{ss}, \qquad e^{(k,s)}=E_{ks}, \qquad
f^{(k,s)}=E_{sk}, \qquad k\lnu s .
\end{equation}
Then the scattering matrix $T_\nu (\lambda ) $ will be a product
of mutually commuting matrices $T_\nu ^{(k,s)} $ of the form:
\begin{equation}\label{eq:T-km}
T_\nu ^{(k,s)} = \openone + (a_{\nu ;ks}^{+}(\lambda ) -1 ) E_{kk} +
(a_{\nu ;ks}^{-}(\lambda ) -1 ) E_{ss} - b_{\nu ;ks}^{-}(\lambda )E_{ks}
+ b_{\nu ;ks}^{+}(\lambda )E_{sk},
\end{equation}
where $k\lnu s $, with only 4 non-trivial matrix elements, just
like the ZS (or AKNS) system.

The $\bbbz_n $-symmetry imposes the following constraints on the FAS and
on the scattering matrix and its factors:
\begin{subequations}\label{eq:Z_n-cons}
\begin{eqnarray}\label{eq:Z_n-cons.a}
&& C_0\xi^\nu (x,\lambda \omega ) C_0^{-1} = \xi^{\nu -2} (x,\lambda ) ,
\qquad C_0 T_\nu (\lambda \omega ) C_0^{-1} = T_{\nu -2}(\lambda ), \\
\label{eq:Z_n-cons.b}
&& C_0 S^\pm_\nu (\lambda \omega ) C_0^{-1} = S^\pm_{\nu -2}(\lambda ),
\qquad C_0 D^\pm_\nu (\lambda \omega ) C_0^{-1} = D^\pm_{\nu -2}(\lambda
),
\end{eqnarray}
\end{subequations}
where the index $\nu -2 $ should be taken modulo $2n $.
Consequently we can view as independent only the data on two of the
rays, e.g. on $l_1 $ and $l_{2n}\equiv l_0 $; all the rest will be
recovered from (\ref{eq:Z_n-cons}).

If in addition we impose the $\bbbz_2 $-symmetry (\ref{eq:Z_2}),
(\ref{eq:Z-N}) with $\epsilon =-1 $  then we will have also $a_k=i\omega
^{k-1/2} $ and:
\begin{equation}\label{eq:Z_2-cons}
B(\xi^\nu (x,-\lambda ^*))^\dag B^{-1} = (\xi^{n+1-\nu}(x,\lambda))^{-1},
\qquad B(S_\nu ^\pm (\lambda ^*))B^{-1} = (S_{n+1-\nu}^{\mp}(\lambda
))^{-1},
\end{equation}
and analogous relations for $T_\nu ^\pm(\lambda )
$ and $D_\nu ^\pm(\lambda ) $. Another interesting subcase takes place for
even values of $n $ and $\bbbz_2 $-reduction (\ref{eq:Z_2}),
(\ref{eq:Z-N}) with $\epsilon =1 $; then $a_k=\omega ^{k-1/2} $ and
the FAS satisfy:
\begin{equation}\label{eq:Z_2-cons'}
B(\xi^\nu (x,\lambda ^*))^\dag B^{-1} = (\xi^{2n+1-\nu}(x,\lambda))^{-1},
\qquad B(S_\nu ^\pm (\lambda ^*))B^{-1} = (S_{2n+1-\nu}^{\mp}(\lambda
))^{-1},
\end{equation}
In both cases the rays $l_\nu  $ are defined by (\ref{eq:l-nu})
with $\phi_0 =0$. The pairs of indices $\{k_\nu , m_\nu \} $
specifying the imbeddings of the $sl(2) $-subalgebras related to
the ray $l_\nu  $ are defined as follows:
\begin{eqnarray}\label{eq:sl2-lnu}
&& \mbox{a) for $\epsilon =1 $} \qquad k_\nu +m_\nu = \left[{n
\over 2
}\right] + 2 - \nu  (\mod n), \nonumber\\
&& \mbox{b) for $\epsilon =-1 $} \qquad k_\nu +m_\nu = 2 - \nu
(\mod n),
\end{eqnarray}

One can prove also that $D_\nu ^+(\lambda ) $ (resp. $D_\nu
^-(\lambda ) $) allows analytic extension for $\lambda \in \Omega
_\nu  $ (resp. for $\lambda \in \Omega _{\nu -1} $, compare with
corollary \ref{cor:7.1}. Another important fact is \cite{GeYa}
that  $D_\nu ^+(\lambda ) = D_{\nu +1}^-(\lambda ) $ for all
$\lambda \in \Omega _\nu  $.

\section*{The inverse scattering problem and the Riemann-Hilbert problem.
}\label{sec:ism-rhp}

The next important step is the possibility to reduce the solution of the
ISP for the GZSs to a (local) RHP. Indeed the relation (\ref{eq:14.3}) can
be rewritten as:
\begin{subequations}\label{eq:23.1}
\begin{eqnarray}\label{eq:23.1a}
\xi^+(x,t,\lambda )&=& \xi^-(x,t,\lambda ) G(x,t,\lambda ), \qquad \lambda
\in \bbbr, \\
\label{eq:23.1b}
G(x,t,\lambda )&=& e^{-i(\lambda Jx-f(\lambda )t)} G_0(\lambda )
e^{i(\lambda Jx-f(\lambda )t)}, \\
\label{eq:23.1c}
G_0(\lambda ) &=& \left. \hat{S}^-(\lambda)S^+(\lambda ) \right|_{t=0};
\end{eqnarray}
\end{subequations}
in other words the sewing function $G(x,t,\lambda ) $ satisfies the
equations:
\begin{eqnarray}\label{eq:23.2}
i {dG  \over dx } -\lambda [J,G(x,t,\lambda )]=0, \qquad
i {dG  \over dt } +[f(\lambda) ,G(x,t,\lambda )]=0,
\end{eqnarray}
Here $f(\lambda) \in {\cal h}$ determines the dispersion law of
the NLEE. Together with
\begin{equation}\label{eq:23.2n}
\lim_{\lambda \to\infty } \xi^\pm(x,\lambda ) =\openone ,
\end{equation}
eq. (\ref{eq:23.1}) is known as the RHP with canonical normalization.

\begin{theorem}[\cite{Sha}]\label{th:Sha} Let $\xi^+(x,t,\lambda ) $ and
$\xi^-(x,t,\lambda ) $ be solutions to the RHP (\ref{eq:23.1}),
(\ref{eq:23.2n}) allowing analytic extension in $\lambda  $ for $\lambda
\in \bbbc_\pm $ respectively. Then $\chi^\pm(x,t,\lambda )
=\xi^\pm(x,t,\lambda )e^{i\lambda Jx} $ are fundamental analytic solutions
of both operators $L $ and $M $, i.e. satisfy  eqs. (\ref{eq:2.2}),
(\ref{eq:2.3}) with
\begin{equation}\label{eq:23.3}
q(x,t) = \lim_{\lambda  \to\infty } \lambda \left(J - \xi^\pm(x,t,\lambda
) J \hat{\xi}^\pm(x,t,\lambda )\right).
\end{equation}
\end{theorem}

\begin{proof}
Let us assume that $\xi^\pm(x,t,\lambda ) $ are regular solutions to the
RHP and let us introduce the function:
\begin{equation}\label{eq:24.1}
g^\pm(x,t,\lambda ) = i{d\xi^\pm  \over dx } \hat{\xi}^\pm(x,t,\lambda ) +
\lambda \xi^\pm(x,t,\lambda )J \hat{\xi}^\pm(x,t,\lambda ).
\end{equation}
If $\xi^\pm(x,t,\lambda ) $ are regular then neither $\xi^\pm(x,t,\lambda
) $ nor their inverse $\hat{\xi}^\pm(x,t,\lambda ) $ have singularities in
their regions of analyticity $\lambda \in \bbbc_\pm $. Then the functions
$g^\pm(x,t,\lambda ) $ also will be regular for all $\lambda \in
\bbbc_\pm$. Besides:
\begin{equation}\label{eq:24.2}
\lim_{\lambda \to\infty }g^+(x,t,\lambda ) = \lim_{\lambda \to\infty }
g^-(x,t,\lambda ) = \lambda J.
\end{equation}
The crucial step in the proof of \cite{ZaSha} is based on the chain of
relations:
\begin{eqnarray}\label{eq:24.3}
g^+(x,t,\lambda ) &\mathop{=}\limits^{(\ref{eq:23.1})}&
i {d (\xi^-G ) \over dx } \hat{G}\hat{\xi}^-(x,t,\lambda ) + \lambda
\xi^-G J\hat{G}\hat{\xi}^-(x,t,\lambda ) \nonumber\\
&=& i {d\xi^-  \over dx }\hat{\xi}^-(x,t,\lambda ) + \xi^- \left(
i {dG  \over dx }\hat{G} + \lambda GJ\hat{G}(x,t,\lambda )\right)
\hat{\xi}^-(x,t,\lambda ) \nonumber\\
&\mathop{=}\limits^{(\ref{eq:23.2})}&
i {d\xi^-  \over dx }\hat{\xi}^-(x,t,\lambda ) + \xi^- \left(
\lambda [J,G]\hat{G} + \lambda GJ\hat{G}(x,t,\lambda )\right)
\hat{\xi}^-(x,t,\lambda ) \nonumber\\
&=& i {d\xi^-  \over dx }\hat{\xi}^-(x,t,\lambda ) + \lambda  \xi^-
J\hat{\xi}^-(x,t,\lambda ) \nonumber\\
&\equiv& g^-(x,t,\lambda ), \qquad \lambda \in \bbbr.
\end{eqnarray}
Thus we conclude that $g^+(x,t,\lambda )=g^-(x,t,\lambda ) $ is a
function analytic in the whole complex $\lambda  $-plane except in the
vicinity of $\lambda \to\infty  $ where $g^+(x,t,\lambda ) $ tends to
$\lambda J $, (\ref{eq:24.2}). Next from Liouville theorem
we conclude that the difference $g^+(x,t,\lambda ) -\lambda J$ is a
constant with respect to $\lambda  $; if we denote this `constant' by
$-q(x,t) $ we get:
\begin{equation}\label{eq:24.4}
g^+(x,t,\lambda ) -\lambda J = -q(x,t).
\end{equation}
It remains to remember the definition of $g^+(x,t,\lambda ) $
(\ref{eq:24.1}) to find that $\xi^\pm(x,t,\lambda ) $ satisfy
(\ref{eq:N16.1}), i.e. that $\chi ^\pm(x,t,\lambda ) $ is a fundamental
solution to $L $. The relation between $q(x,t) $ and $\xi^\pm(x,t,\lambda
) $ (\ref{eq:23.3}) can be obtained by taking the limit of the
left-hand sides of (\ref{eq:24.4}) for $\lambda \to\infty  $.

Arguments along the same line applied to the functions $h^\pm(x,t,\lambda
) $
\begin{equation}\label{eq:24.5}
h^\pm(x,t,\lambda )= i{d\xi^\pm  \over dt } \hat{\xi}^\pm(x,t,\lambda ) -
\xi^\pm(x,t,\lambda )f(\lambda ) \hat{\xi}^\pm(x,t,\lambda ),
\end{equation}
can be used to prove that $\chi ^\pm(x,t,\lambda ) $ are fundamental
solutions also of the operator $M $; equivalently it satisfies
($V'(x,t,\lambda )=V(x,t,\lambda )-f(\lambda ) $):
\begin{equation}\label{eq:24.6}
i {d\xi^\pm  \over dt } + V'(x,t,\lambda ) \xi^\pm(x,t,\lambda ) +
[f(\lambda ), \xi^\pm(x,t,\lambda )]=0,
\end{equation}
and one finds that $h^+(x,t,\lambda )=h^-(x,t,\lambda ) $ is a function
analytic everywhere in $\bbbc $ except at $\lambda \to\infty  $ where it
has a polynomial behavior of order $N-1 $. Denoting the polynomial as
$V(x,t,\lambda ) $ we derive (\ref{eq:24.5}).

To conclude the proof of the theorem we have to account for
possible zeroes and pole singularities  of $\xi^\pm(x,t,\lambda ) $
at the points ${\cal  Z} $ (\ref{eq:Zero}).  Below we derive the
structure of these singularities which is such that they do not
influence the functions $g^\pm(x,t,\lambda ) $ and $h^\pm(x,t,\lambda ) $.
The theorem is proved.  \end{proof}

The analyticity properties of  $m_k^\pm(\lambda ) $ allow one to
reconstruct  them from the sewing function $G(\lambda ) $ (\ref{eq:23.1c})
and from the locations of their zeroes through (see Appendix~\ref{sec:A2}):
\begin{equation}\label{eq:N23.3}
{\cal  D}_k(\lambda ) = {1 \over 2\pi i } \int_{-\infty }^{\infty } { d\mu
\over \mu -\lambda  } \ln \left\{ \begin{array}{cccc} 1, & 2, & \dots , &
k \\ 1, & 2, & \dots , & k \end{array} \right\}_{G(\mu )} + \sum_{j=1}^{N}
b_{jk} \ln {\lambda -\lambda _j^+  \over  \lambda -\lambda _j^-},
\end{equation}
where
\begin{equation}\label{eq:N23.4}
{\cal  D}_k(\lambda ) = \left\{ \begin{array}{ll} \ln m_k^+(\lambda ),
\qquad \lambda \in \bbbc_+ \\ -\ln m_{n-k}^-(\lambda ),
\qquad \lambda \in \bbbc_- . \end{array} \right.
\end{equation}

One can view ${\cal  D}_k(\lambda ) $ as generating functionals of the
conserved quantities for the related $N $-wave-type equations; the
relevant expressions for them in terms of the scattering data can be
obtained from the right hand sides of (\ref{eq:N23.3}).

Quite analogously we can treat also the CBCs with  $\bbbz_n $-symmetry.
More precisely, we have:
\begin{eqnarray}\label{eq:*-nu}
\xi^\nu (x,t,\lambda ) = \xi^{\nu -1}(x,t,\lambda ) G_\nu (x,t,\lambda ),
\qquad \lambda \in l_\nu ,\\
G_\nu (x,t,\lambda ) = e^{-i\lambda Jx +if(\lambda )t} G_{0,\nu
}(\lambda ) e^{i\lambda Jx-if(\lambda )t},  \qquad G_{0,\nu
}(\lambda )= \left. \hat{S}_\nu ^-(\lambda ) S_\nu ^+(\lambda
)\right|_{t=0} \nonumber
\end{eqnarray}
The collection of all relations (\ref{eq:*-nu}) for $\nu =1,2,\dots,2n $
together with
\begin{equation}\label{eq:*-nu-norm}
\lim_{\lambda \to\infty } \xi^\nu (x,t,\lambda ) = \openone ,
\end{equation}
can be viewed as a local RHP posed on the collection of rays
$\Sigma \equiv \{l_\nu \}_{\nu =1}^{2n} $ with canonical normalization.
Rather straightforwardly we can reformulate the results for the GZSs for
the CBCs and prove that if $\xi^\nu (x,\lambda ) $ is a solution of the
RHP (\ref{eq:*-nu}), (\ref{eq:*-nu-norm}) then $\chi ^\nu (x,\lambda
)=\xi^\nu (x,\lambda ) e^{i\lambda Jx} $ satisfy the CBC with potential
\begin{equation}\label{eq:q-CBC}
q(x,t) = \lim_{\lambda \to\infty } \left( J - \xi^\nu (x,t,\lambda ) J
\hat{\xi}^\nu (x,t,\lambda ) \right).
\end{equation}
We finish this subsection by formulating the dispersion relations
for the functions $\ln m_{\nu ,k}^{+}(\lambda )$ which allows us
to reconstruct them from their analyticity properties:
\begin{equation}\label{eq:dis-rel}
\ln  m_{\nu ,k}^+(\lambda ) = \sum_{\eta =1}^{2n} {(-1)^\eta \over
2\pi i} \int_{l_\nu } {d\mu   \over \mu -\lambda } \ln \left\{
\begin{array}{ccc} 1 & \dots & k \\ 1 & \dots & k  \end{array}
\right\}_{G_\eta (\mu )}^{\eta} + \sum_{\eta =1}^{n}
\sum_{j=1}^{N} b_{kj}^{\eta} \ln { \lambda -\lambda _{j,k}^+\omega
^{\eta } \over \lambda - \lambda _{j,k}^-\omega ^{\eta }  } ,
\end{equation}
where $\lambda \in \Omega_\nu$ and the superscript $\eta  $ in the
integrand shows that we should use the ordering characteristic for
the sector $\Omega_\eta $; $b_{kj}^{\eta} $ are the analogs for
$b_{kj} $ (\ref{eq:N25.6}) in $\Omega _\eta$.

Both dispersion relations (\ref{eq:N23.3}) and  (\ref{eq:dis-rel})
can be used to derive the so-called trace identities (see
\cite{ZMNP,FaTa}) for the GZSs and CBCs respectively. Indeed,
${\cal D}_k (\lambda)$ and $\ln m_{\nu,k}^+ (\lambda)$ allow
asymptotic expansions
\begin{equation}\label{eq:as-exp}
{\cal D}_k (\lambda) = \sum_{s=1}^{\infty} {\cal D}_k^{(s)}
\lambda^{-s}, \qquad \ln m_{\nu,k}^+ (\lambda) =
\sum_{s=1}^{\infty} M_{\nu,k}^{(s)} \lambda^{-s}.
\end{equation}
The expansion coefficients ${\cal D}_k^{(s)}$ and
$M_{\nu,k}^{(s)}$ are local integrals of motion, i.e. their
densities depend only on $q(x,t)$ and its derivatives with respect
to $x$. Their explicit calculation is done via recurrent
procedure. We illustrate it by the two first integrals of motion
of the $\bbbz_n$-NLS equation (\ref{eq:3.3}):
\begin{eqnarray}\label{eq:2.51'}
M_{1,1}^{(1)} &=& {1 \over 2\omega } \int_{-\infty}^{\infty} dx \,
\sum_{p=1}^n \psi_p \psi_{n-p} (x,t), \\ \label{eq:2.51''}
M_{1,1}^{(2)} &=& {1 \over 2\omega ^2} \int_{-\infty}^{\infty} dx
\,\left\{ \sum_{p=1}^n i \,\cotan \left({\pi p \over n}\right)
\left( { d\psi_{p} \over dx} \psi_{n-p} - \psi_{p} {d\psi_{n-p} \over dx}
\right) \right.  \\ && \qquad \left. - {2\over 3} \sum_{p+k+l=n} \psi_{p}
\psi_{k}\psi_{l}(x,t) \right\} , \nonumber
\end{eqnarray}

One can also expand the right hand sides of the dispersion
relations (\ref{eq:N23.3}) and  (\ref{eq:dis-rel}) over the
inverse powers of $\lambda$ which allows to express ${\cal
D}_k^{(s)}$ and $M_{\nu,k}^{(s)}$ also in terms of the scattering
data of GZSs and CBCs.

\section*{The dressing Zakharov-Shabat method }\label{sec:dres}

One of the most fruitful ideas for the explicit constructing of
the NLEE's solutions is based on the possibility starting from a
given regular solutions $\xi_0^\pm(x,t,\lambda ) $ to the RHP to
construct new singular solutions $\xi^{\pm}(x,t,\lambda ) $
having zeroes and pole singularities at the prescribed points
$\lambda _j^\pm \in \bbbc_\pm $. The structure of these
singularities are determined by the dressing factor
$u_j(x,t,\lambda ) $:
\begin{equation}\label{eq:26.1}
\xi^{\pm}(x,t,\lambda ) = u_j(x,t,\lambda ) \xi_0^{\pm}(x,t,\lambda )
u_{j,-}^{-1}(\lambda ),
\end{equation}
which for the $SL(n) $-group has a simple fraction-linear dependence on $
\lambda  $:
\begin{eqnarray}\label{eq:26.2}
u_j(x,t,\lambda ) &=& \openone + (c_j(\lambda ) -1)P_j(x,t), \qquad
c_j(\lambda ) = {\lambda  -\lambda _j^+ \over \lambda  -\lambda _j^- }, \\
u_{j,-}^{-1} &=& \lim_{x\to -\infty } u_j(x,t,\lambda ).
\end{eqnarray}
Here $P_j(x,t) $ is a projector $P_j^2=P_j $ which for simplicity
is chosen to be of rank $1$; then it can be written down as:
\begin{equation}\label{eq:26.3}
P_j(x) = {|n_{j}\rangle\langle m_{j}| \over \langle m_{j}| n_{j}\rangle},
\end{equation}
where the bra- and ket- eigenvectors $\langle m_{j}| $ and $|n_{j}\rangle
$ are the `left' and `right' eigenvectors of the projector.

{}From (\ref{eq:26.1}) there follows that the dressing factor
$u(x,t,\lambda )$ satisfies the equation:
\begin{equation}\label{eq:u-eq}
i {du  \over dx } + q(x,t) u(x,t,\lambda ) - u(x,t,\lambda ) q_0(x,t) -
\lambda [J,u(x,t,\lambda )] =0.
\end{equation}

The main advantage of the dressing method is in the fact that one can
determine the $x $ and $t $-dependence of $\langle m_{j}| $ and
$|n_{j}\rangle $ through the regular solution $\chi_0^{\pm}(x,t,\lambda )
$ as follows:
\begin{eqnarray}\label{eq:27.1'}
\qquad |n_{j}\rangle =\chi_{0j}^{+}(x,t) |n_{j}^0\rangle , \qquad
\langle m_{j}| = \langle m_{j}^0 |\hat{\chi}_{0j}^{-}(x,t), \qquad
\chi_{0j}^{\pm}(x,t)=\chi_0^{\pm}(x,t,\lambda_j^\pm )
\end{eqnarray}
or equivalently these vectors are solutions to the equations:
\begin{eqnarray}\label{eq:27.1}
i {d|n_{j}\rangle  \over dx } + U^{(0)}(x,t,\lambda _j^+)|n_{j}
\rangle =0, &\quad &i {d|n_j\rangle \over dt } + V^{(0)}(x,t,\lambda_j^+)
|n_{j}\rangle =0,\\
i {d\langle m_{j} |\over dx } - \langle m_{j}| U^{(0)}(x,t,\lambda
_j^-)=0, &\quad & i {d\langle m_{j} |\over dt } -\langle m_{j}|
V^{(0)}(x,t,\lambda_j^-)=0,\\
U^{(0)}(x,t,\lambda )=q_0(x,t) - \lambda J, &\quad &
V^{(0)}(x,t,\lambda )=\left. V(x,t,\lambda )\right|_{q=q_0}.
\end{eqnarray}
Here  $q_0(x,t) $ is  the potential corresponding to the regular
solutions $\chi_0^\pm(x,t,\lambda ) $ to the RHP and
$V^{(0)}(x,t,\lambda ) $ is obtained from $V(x,t,\lambda ) $ (see
(\ref{eq:7.8}), (\ref{eq:7.8'})) replacing $q(x,t) $ by $q_0(x,t)
$. This construction is well defined also in the case when $\chi
_0^\pm(x,\lambda ) $ are singular solutions to the RHP, provided
they are regular for $\lambda =\lambda _j^\pm $.

If  $q(x,t) $ is the potential corresponding to the singular
solution $\chi^{\pm}(x,t,\lambda )  $ then:
\begin{eqnarray}\label{eq:27.2}
q(x,t) &=& q_0(x,t) + \lim_{\lambda \to\infty } \lambda (J -
u_j(x,t,\lambda ) J\hat{u}_j(x,t,\lambda )) \nonumber\\
&=& q_0(x,t) - (\lambda _j^+ - \lambda _j^-) [J,P_j(x,t)].
\end{eqnarray}

Thus starting from a given regular solution of the RHP (and
related solution $q_0(x,t) $ to the NLEE) we can construct a
singular solution to the RHP and a new solution $q(x,t) $ of the
NLEE depending on the $\lambda _j^\pm $ and on the eigenvectors of
$P_j(x) $. If we start from the trivial solution $q_0(x,t)=0 $ of
the NLEE then we will get the one-soliton solution of the NLEE.
Repeating the procedure $N $ times we can get the $N $-soliton
solution of the NLEE.

With the explicit formulae for $P_j(x) $ and using (\ref{eq:26.1})
we can establish the relationship between the scattering data of
the regular RHP and the corresponding singular one. The dressing
factor $u_j(x,\lambda ) $ is determined by the constant vectors
$\langle m_{j}^0| $ and $|n_{j}^0\rangle $ can not be quite
arbitrary.  The condition that $q(x) $ vanishes for $x\to\pm\infty
$ requires that if $(n_j^0)_s =0$ for all $1\leq s<I_j $ and
$F_j<s\leq n $ then also $(m_j^0)_s =0$ for all $1\leq s<I_1 $ and
$F_1<s\leq n $. Thus we derive that:
\begin{equation}\label{eq:lim-Pj}
\lim_{x\to\infty } P_j(x) = E_{I_jI_j}, \qquad \lim_{x\to -\infty } P_j(x)
= E_{F_jF_j},
\end{equation}
and therefore
\begin{equation}\label{eq:uj-pm}
u_{j,+}(\lambda )= \openone + (c_j(\lambda )-1) E_{I_jI_j}, \qquad
u_{j,-}(\lambda )= \openone + (c_j(\lambda )-1) E_{F_jF_j}.
\end{equation}
The interrelation between the Gauss factors of the corresponding
scattering matrices are:
\begin{eqnarray}\label{eq:Spm}
&& S^{\pm}(\lambda ) = u_{j,-}(\lambda ) S_{0}^{\pm}(\lambda )
u_{j,-}^{-1}(\lambda ), \qquad  T^{\pm}(\lambda ) = u_{j,+}(\lambda )
T_{0}^{\pm}(\lambda ) u_{j,+}^{-1}(\lambda ),
\end{eqnarray}
and
\begin{equation}\label{eq:Dpm}
D^{\pm}(\lambda ) = u_{j,+}(\lambda ) D_{0}^{\pm}(\lambda )
u_{j,-}^{-1}(\lambda ).
\end{equation}
Comparing these last relations with (\ref{eq:15.2}) we find for the
principal minors of $T(\lambda ) $ and $T_0(\lambda ) $:
\begin{subequations}\label{eq:ms-pm}
\begin{eqnarray}\label{eq:ms-pm-a}
&& \qquad  m_{s}^{+}(\lambda ) = {\lambda -\lambda _j^+ \over \lambda
-\lambda _j^- } m_{0,s}^{+}(\lambda ) , \quad \mbox{for} \quad I_j \leq s
<F_j, \quad \lambda \in \bbbc_+\cup\bbbr, \\
\label{eq:ms-pm-b}
&& \qquad  m_{s}^{-}(\lambda ) = {\lambda -\lambda _j^- \over \lambda
-\lambda _j^+ } m_{0,s}^{+}(\lambda ) , \quad \mbox{for} \quad n-F_j < s
\leq n-I_j, \quad \lambda \in \bbbc_-\cup\bbbr,
\end{eqnarray}
\end{subequations}
and $m_{s}^{\pm}(\lambda )= m_{0,s}^{\pm}(\lambda ) $ for the other
values of $s $. Thus the result of the dressing is that the string of
upper principle minors $m_{s}^{+}(\lambda ) $, $I_j\leq s<F_j $ acquire
simple zero at $\lambda =\lambda _j^+ $ while the string of
lower principle minors $m_{s}^{-}(\lambda ) $, $n-F_j< s\leq n-I_j $
acquire simple zero at $\lambda =\lambda _j^- $.

Obviously if we impose on $L(\lambda ) $ the $\bbbz_2 $-reduction then we
should restrict also the dressing factor by:
\begin{equation}\label{eq:Z_2-u}
B\left( u(x,t,\epsilon \lambda ^*)\right)^\dag B^{-1} = u(x,t,\lambda ).
\end{equation}
The ansatz (\ref{eq:26.2}) satisfies (\ref{eq:Z_2-u}) if $\lambda _j^- =
\epsilon (\lambda _j^+)^* $ and the vectors $|n_{0j}\rangle  $, $\langle
m_{0j}| $ are related by:
\begin{equation}\label{eq:Z_2-nm}
\langle m_{0j}| = B |n_{0j}^\dag \rangle .
\end{equation}
If we impose the $\bbbz_n $-reduction (\ref{eq:Z_n}) then $u(x,t,\lambda )
$ must satisfy:
\begin{equation}\label{eq:Z_n-u}
C_0 u(x,t,\omega \lambda )C_0^{-1} = u(x,t,\lambda ).
\end{equation}
Such conditions require generalizations of the ansatz (\ref{eq:26.2})
\cite{Mikh}:
\begin{equation}\label{eq:26.2-n}
u_j(x,t,\lambda ) = \openone + \sum_{s=0}^{n-1} \left( c_j(\omega
^s\lambda ) - 1 \right) C_0^{s}P_j(x) C_0^{-s}.
\end{equation}
A slightly different approach treating also multi-soliton
solutions of the $\bbbz_n$-symmetric NLEE is given in
\cite{BeCo1}.

Up to now we dealt with the algebra $\fr{g}\simeq sl(n,\bbbc) $.
Treating the other simple Lie algebras (orthogonal or symplectic)
needs additional care especially in constructing the dressing
factors \cite{ZaMi,GGKI}.

In fact $u_j(x,\lambda ) $ (\ref{eq:26.2}) must be an element of
the corresponding group. From the ansatz (\ref{eq:26.2}) it
follows that $u_j(x,\lambda ) $ belongs to $GL(n,\bbbc) $, but one
can always multiply $u(x,\lambda ) $ by an appropriate $x $- and
$t $-independent scalar and to adjust its determinant to 1. Such a
multiplication goes through the whole scheme outlined above but is
adequate only for the $sl(n,\bbbc) $ case.  However the ansatz
(\ref{eq:26.2}) can not be used, e.g. for the case $so(n,\bbbc) $.
The adequate ansatz is formulated below \cite{GGKI}.

\begin{theorem}\label{th:1}
Let $\fr{g}\sim {\bf B}_r $ or ${\bf D}_r $  and let the dressing factor $
u(x,\lambda ) $ be of the form:
\begin{equation}\label{eq:u-lam}
u_j(x,\lambda )= \openone + (c_j(\lambda ) -1)P_j(x) + (c_j^{-1}(\lambda )
-1) P_{-j}(x), \qquad P_{-j} =S_0P_j^TS_0^{-1},
\end{equation}
where $S_0 $ is introduced in (\ref{eq:def-S0}) and $P_j(x) $ is a rank 1
projector (\ref{eq:26.3}). Let the constant vectors $|n_0\rangle  $ and
$\langle m_0| $ satisfy the condition
\begin{equation}\label{eq:m-m}
\langle m_0 | S|m_0 \rangle = \langle n_0 | S|n_0 \rangle =0.
\end{equation}
Then $u_j(x,\lambda ) $ (\ref{eq:u-lam}) satisfies the equation
(\ref{eq:u-eq}) with a potential
\begin{equation}\label{eq:40.6}
q(x) = q_{0}(x) - (\lambda _j^+ - \lambda _j^-) [J,p_j(x)], \qquad
p_j(x) = P_j(x) - P_{-j}(x).
\end{equation}
\end{theorem}

\begin{proof}
Due to the fact that $\chi_0 ^\pm (x,\lambda ) $ take values in the
corresponding orthogonal group we find that from (\ref{eq:m-m}) it follows
$ \langle m | S|m \rangle =0$, $\langle m | JS|m \rangle =0$ and analogous
relations for the vector $|n\rangle  $.  As a result we get that
\begin{equation}\label{eq:p1p-1}
P_j(x) P_{-j}(x) =P_{-j}(x) P_j(x) =0, \qquad
P_j(x)J P_{-j}(x) =P_{-j}(x)J P_j(x) =0.
\end{equation}

Let us now insert (\ref{eq:u-lam}) into (\ref{eq:u-eq}) and take the
limit of the r.h.side of (\ref{eq:u-eq}) for $\lambda \to \infty $. This
immediately gives eq. (\ref{eq:40.6}). In order that  Eq.~(\ref{eq:u-eq})
be satisfied identically with respect to $\lambda $ we have to put to 0
also the residues of its r.h.side at $\lambda \to \lambda_j^+ $ and
$\lambda \to \lambda_j^- $. This gives us the following system of equation
for the projectors $P_j(x) $ and $P_{-j}(x) $:
\begin{eqnarray}\label{eq:p1-p1}
i {d P_j  \over dx } + q(x) P_j(x) - P_j(x) q_{0}(x) - \lambda
_j^- [J, P_j(x)]=0, \\
i {d P_{-j} \over dx } + q(x) P_{-j}(x) - P_{-j}(x) q_{0}(x) -
\lambda _j^+ [J, P_{-j}(x)]=0,
\end{eqnarray}
where we have to keep in mind that $q $ is given by (\ref{eq:40.6}).
Taking into account (\ref{eq:p1p-1}) and the relation between $P_j(x) $ and
$P_{-j}(x) $ eq. (\ref{eq:p1-p1}) reduces to:
\begin{equation}\label{eq:P1}
i {d P_j \over dx } + [q_{0}(x), P_j(x)] + \lambda_j^-  P_j(x)J  -
\lambda_j^+ J P_j(x) -(\lambda_j^- -\lambda _j^+) P_j(x) JP_j(x)=0.
\end{equation}
One can check by a direct calculation that (\ref{eq:26.3}) satisfies
identically (\ref{eq:P1}). The theorem is proved.

\end{proof}

\specialsection{The resolvent and spectral properties of
GZSs and CBCs}\label{sec:3}

The FAS $\chi ^\pm(x,\lambda ) $ of $L(\lambda ) $ allows one to construct
the resolvent of the operator $L $ and then to investigate its spectral
properties. By resolvent of $L(\lambda ) $ we understand the integral
operator $R(\lambda ) $ with kernel $R(x,y,\lambda ) $ which satisfies
\begin{equation}\label{eq:R1.1}
L(\lambda )(R(\lambda )f)(x)=f(x),
\end{equation}
where $f(x) $ is an $n $-component vector function in $\bbbc^n $ with
bounded norm, i.e. $\int_{-\infty }^{\infty } dy (f^T(y) f(y)) <\infty $.

{}From the general theory of linear operators \cite{AG,DS,M24} we
know that the point $\lambda $ in the complex $\lambda  $-plane is
a regular point if $R(\lambda ) $ is a bounded integral operator.
In each connected subset of regular points $R(\lambda ) $ is
analytic in $\lambda  $.

The points $\lambda  $ which are not regular constitute the spectrum of
$L(\lambda ) $. Roughly speaking the spectrum of $L(\lambda ) $ consist of
two types of points:

\begin{itemize}

\item i) the continuous spectrum of $L(\lambda ) $ consists of all points
$\lambda  $ for which $R(\lambda ) $ is an unbounded integral operator;

\item ii) the discrete spectrum of $L(\lambda ) $ consists of all points
$\lambda  $ for which $R(\lambda ) $ develops pole singularities.

\end{itemize}

Let us now show how the resolvent $R(\lambda ) $ can be expressed through
the FAS of $L(\lambda ) $. Indeed, if we write down $R(\lambda ) $ in the
form:
\begin{equation}\label{eq:R1.2}
R(\lambda ) f(x) = \int_{-\infty }^{\infty } R(x,y,\lambda ) f(y),
\end{equation}
the kernel $R(x,y,\lambda ) $ of the resolvent is given by:
\begin{equation}\label{eq:R2.1}
R(x,y,\lambda ) = \left\{ \begin{array}{ll}
R^+(x,y,\lambda ) & \mbox{for\;} \lambda \in \bbbc^+, \\
R^-(x,y,\lambda ) & \mbox{for\;} \lambda \in \bbbc^-,
\end{array} \right.
\end{equation}
where
\begin{eqnarray}\label{eq:R2.2}
R^\pm (x,y,\lambda )&=&\pm i\chi ^\pm(x,\lambda ) \Theta ^\pm (x-y)
\hat{\chi }^\pm (y,\lambda ), \\
\Theta ^\pm(z) &=& \theta (\mp z) \Pi_0 - \theta (\pm z) (\openone -\Pi_0),
\qquad \Pi_0 = \sum_{s=1}^{k_0} E_{ss}, \nonumber
\end{eqnarray}
where $k_0 $ is the number of positive eigenvalues of $J $; namely:
\begin{equation}\label{eq:R2.3}
a_1>a_2 >\dots >a_{k_0}>0>a_{k_0+1} > \dots >a_n.
\end{equation}
Due to the condition $\tr J = \sum_{s=1}^{n}a_s =0 $, $k_0 $ is
fixed up uniquely.

The next theorem establishes that $R (x,y,\lambda ) $ is indeed the kernel
of the resolvent of $L(\lambda ) $.

\begin{theorem}\label{th:R2}
Let $q(x) $ satisfy conditions (C.1) and (C.2) and let $\lambda _j^\pm $
be the simple zeroes of the minors $m_k^\pm(\lambda ) $. Then
\begin{enumerate}

\item $R^\pm (x,y,\lambda ) $ is an analytic function of $\lambda  $ for
$\lambda \in \bbbc_\pm $ having pole singularities at $\lambda _j^\pm \in
\bbbc_\pm $;

\item $R^\pm (x,y,\lambda ) $ is a kernel of a bounded integral operator
for $\im \lambda \neq 0 $;

\item $R (x,y,\lambda ) $ is uniformly bounded function for $\lambda
\in\bbbr $ and provides a kernel of an unbounded integral operator;

\item $R^\pm (x,y,\lambda ) $ satisfy the equation:
\begin{equation}\label{eq:R3.1}
L(\lambda ) R^\pm (x,y,\lambda )=\openone \delta (x-y).
\end{equation}
\end{enumerate}
\end{theorem}

\begin{proof}[Idea of the proof] {}

\begin{enumerate}

\item is obvious from the fact that $\chi ^\pm(x,\lambda ) $ are the FAS
of $L(\lambda ) $;

\item Assume that $\im \lambda >0 $ and consider the asymptotic behavior
of $R^+ (x,y,\lambda ) $ for $x,y\to\infty  $. From equations
(\ref{eq:s6.1}) we find that
\begin{eqnarray}\label{eq:R3.2}
R_{ij}^+ (x,y,\lambda ) &=& \sum_{p=1}^{n} \xi^+_{ip}(x,\lambda )
e^{-i\lambda a_p(x-y)} \Theta^+_{pp}(x-y) \hat{\xi}^+_{pj}(y,\lambda ) \\
\nonumber
\end{eqnarray}

Due to the fact that $\chi ^+(x,\lambda ) $ has triangular asymptotics
for $x\to\infty  $ and $\lambda \in\bbbc_+ $ and for the correct choice of
$\Theta^+(x-y) $ (\ref{eq:R2.2}) we check that the right hand side of
(\ref{eq:R3.2}) falls off exponentially for $x\to\infty  $ and arbitrary
choice of $y $. All other possibilities are treated analogously.

\item For $\lambda \in\bbbr$ the arguments of 2) can not be applied
because the exponentials in the right hand side of (\ref{eq:R3.2})
$\im \lambda =0  $ only oscillate. Thus we conclude that
$R^\pm(x,y,\lambda ) $ for $\lambda \in\bbbr $ is only a bounded
function and thus the corresponding operator $R(\lambda ) $ is an
unbounded integral operator.

\item The proof of eq. (\ref{eq:R3.1}) follows from the fact that
$L(\lambda )\chi ^+(x,\lambda )=0 $ and
\begin{equation}\label{eq:R4.1}
{d\Theta^\pm (x-y)  \over dx } = \mp \openone \delta (x-y).
\end{equation}
\end{enumerate}
\end{proof}

\begin{proposition}\label{pro:ds}
Let $q(x) $ satisfy the conditions (C.1) and (C.2), let ${\cal
Z}\equiv \{\lambda _j^\pm,j=1,\dots,N\} $ be the set of simple zeroes
(\ref{eq:N25.2}) of the minors $m_s^\pm(\lambda ) $ and let $I_j\leq k_0
<F_j $ for all $j=1,\dots,N $.  Then the kernel of the resolvent
$R^+(x,y,\lambda ) $ (resp.  $R^-(x,y,\lambda ) $) has simple poles for
$\lambda =\lambda _j^+ $ (resp.  for $\lambda =\lambda _j^- $) with
residues given by:
\begin{subequations}\label{eq:res}
\begin{eqnarray}\label{eq:res0}
&&\res_{\lambda =\lambda _j^\pm} R^\pm(x,y,\lambda ) = \mp 2i\nu _j
\left| {\bf n}_{j}^\pm(x) \right\rangle \left\langle {\bf m}_j^\pm(y)
\right|, \\
\label{eq:res1}
&& \left| {\bf n}_{j}^+(x) \right\rangle = (\openone -P_j(x))
\chi_{0,j}^{+}(x) \Pi_0 |n_{0,j}\rangle , \qquad
\left\langle {\bf m}_j^+(y)\right| = {\langle m_j(y)| \over \langle
m_j(y)| n_j(y) \rangle},  \\
\label{eq:res2}
&& \left| {\bf n}_{j}^-(x) \right\rangle = { | n_{j}(x)\rangle
\over \langle m_j(x)| n_j(x) \rangle  }, \qquad  \left\langle
{\bf m}_j^-(y) \right| = \langle m_{0,j}| \Pi_0 \hat{\chi} _{0,j}^{-}(y)
(\openone -P_1(y)),
\end{eqnarray}
\end{subequations}
where $\lambda _j^\pm=\mu _j\pm i\nu _j $ and $\chi_{0,j}^\pm(x) = \chi
_0^\pm(x,\lambda _j^\pm) $ are the FAS corresponding to the potential $q_0
$ satisfying (C.1) and (C.2) and whose set of simple zeroes is ${\cal
Z}_0\equiv {\cal Z}\backslash \{\lambda _j^+,\lambda _j^-\} $.
\end{proposition}

\begin{proof}
Let $\chi _0^\pm(x,\lambda ) $ be the FAS of $L_0(\lambda ) $ with
potential $q_0(x) $; then $\chi _0^\pm(x,\lambda ) $ are regular for
$\lambda =\lambda _j^\pm $. Now we apply the dressing method choosing
$\lambda _j^\pm $ as the locations of the singularities and construct the
projector $P_j(x) $ using the constant vectors $|n_{0,j}\rangle  $ and
$\langle m_{0,j}| $. The normalizing factor $u_{j,-}^{-1}(\lambda ) $ in
the right hand side of (\ref{eq:26.1}) is a diagonal matrix that commutes
with $\Pi_0 $. Then we insert $\chi ^\pm(x,\lambda )=u_j(x,\lambda )\chi
_0^\pm(x,\lambda ) $ in (\ref{eq:R2.2}) and note that the
pole singularity of $R^+(x,y,\lambda ) $ at $\lambda =\lambda _j^+ $
(resp. $R^-(x,y,\lambda ) $ at $\lambda =\lambda _j^- $) can come up only
from the factor $u_j^{-1}(y,\lambda ) $ (resp. $u(x,\lambda ) $). To
derive the expressions in (\ref{eq:res}) one needs the explicit form of
the projectors $P_j(x) $ and $P_j(y) $ (\ref{eq:26.3}) and
(\ref{eq:27.1'}).

The right hand sides of (\ref{eq:res}) do not vanish if the following
conditions
\begin{equation}\label{eq:P_0-n}
\begin{split}
\Pi_0 |n_{0,j}\rangle \neq |n_{0,j}\rangle,  \qquad \text{or} \qquad
\Pi_0 |n_{0,j}\rangle \neq 0, \\
\langle m_{0,j}|\Pi_0 \neq \langle m_{0,j}|, \qquad \text{or} \qquad
\langle m_{0,j}|\Pi_0 \neq 0.
\end{split}
\end{equation}
hold. In other words if (\ref{eq:P_0-n}) hold then the residues
(\ref{eq:res}) do not vanish, $R^\pm(x,y,\lambda ) $ have simple poles
at $\lambda =\lambda _j^\pm $ and by definition $\lambda _j^\pm $ are
discrete eigenvalues of $L(\lambda ) $. Eq. (\ref{eq:P_0-n}) is
equivalent to the condition $I_j\leq k_0<F_j $.  Indeed  violating this
condition we get either $(\openone -\Pi_0) |n_{0,j}\rangle =0 $ or $\Pi_0
|n_{0,j}\rangle =0 $ and as a result -- vanishing right hand sides in
(\ref{eq:res}).

To finish the proof one must check that from the definitions
(\ref{eq:res1}) the relations (\ref{eq:ms-pm}) follow. Besides
$|{\bf n}_j^\pm\rangle  $ and $\langle {\bf m}_j^\pm| $ satisfy:
\begin{equation}\label{eq:nj-mj}
i {d|{\bf n}_j^\pm\rangle   \over dx } + (q(x) - \lambda _j^\pm J)
|{\bf n}_j^\pm\rangle =0, \qquad
i{d\langle {\bf m}_j^\pm|   \over  dx} - \langle {\bf m}_j^\pm|  (q(x)
-\lambda _j^\pm ) =0,
\end{equation}
where $q(x) $ is given by (\ref{eq:27.2}).
\end{proof}

\begin{corollary}
\label{cor:3.2}
The discrete spectrum of the Lax operator (\ref{eq:2.2}) consists
of the zeroes of the principal minors $m_j^+(\lambda)$ for
$\lambda\in\Bbb{C}_+$ and $m_j^-(\lambda)$ for $\lambda\in\Bbb{C}_-$
provided the conditions (\ref{eq:P_0-n}) are satisfied.
\end{corollary}

Now we can derive the completeness relation for the eigenfunctions of the
Lax operator (\ref{eq:2.2}) by applying the contour  integration method
(see e.g. \cite{GeKu1,GeKu,AKNS}) to the integral:
\begin{equation}\label{eq:3.38}
\cal{J}(x,y)={1\over 2\pi i}\oint_{\gamma _+} d\lambda
R^{+}(x,y,\lambda)- {1\over 2\pi i}\oint_{\gamma _-} d\lambda
R^{-}(x,y,\lambda),
\end{equation}
where the contours $\gamma _\pm$ are shown on the Figure \ref{fig:1}.
Skipping the details we get:

\begin{figure}[tb]
%\blankbox{.6\columnwidth}{5pc}
%\vspace{4in}
%\psfigfile=figg.ps
\includegraphics*{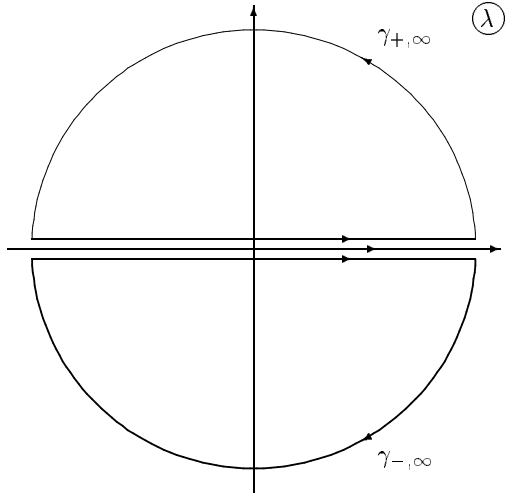}
\caption{The contours $\gamma _\pm =\bbbr\cup\gamma_{\pm\infty }$.}
\label{fig:1}

\end{figure}

\begin{equation}\label{eq:3.44}
\begin{split}
\delta(x-y)  \sum_{s=1}^{n} {1  \over a_s } E_{ss} &= {1\over
2\pi}\int_{-\infty}^\infty d\lambda\left\{ \sum_{s=1}^{k_0}
|\chi^{[s]+}(x,\lambda)\rangle \langle \hat\chi^{[s]+}(y,\lambda)|
\right. \\ & \left. - \sum_{s=k_0+1}^{n}
|\chi^{[s]-}(x,\lambda)\rangle \langle
\hat\chi^{[s]-}(y,\lambda)|\right\} \\ &+ 2i \sum_{j=1}^N \nu _j
\left\{ \left| {\bf n}_j^+(x) \right\rangle \left\langle {\bf
m}_j^+(y) \right| -\left| {\bf n}_j^-(x) \right\rangle
\left\langle {\bf m}_j^-(y) \right| \right\}.
\end{split}
\end{equation}

This relation (\ref{eq:3.44}) allows one to expand any vector-function
$|z(x)\rangle\in \bbbc^{n}$ over the eigenfunctions of the system
(\ref{eq:2.2}).  Indeed, let us multiply (\ref{eq:3.44}) on the right by
$J\mid z(y)\rangle$ and integrate over $y$. This gives:
\begin{eqnarray}\label{eq:3.45}
\mid z(x)\rangle &= {i\over 2\pi}\int_{-\infty}^\infty
d\lambda\left\{ \sum_{s=1}^{k_0}|\chi^{[s]+}(x,\lambda)\rangle
\cdot\zeta_{s}^+(\lambda)
-\sum_{s=k_0+1}^{n}\chi^{[s]-}(x,\lambda) \rangle\cdot
\zeta_{s}^-(\lambda)\right\} \nonumber\\
&+ \sum_{j=1}^{N}\nu _j\left( \left| {\bf n}_j^+(x) \right\rangle \zeta
_j^+ - \left| {\bf n}_j^-(x) \right\rangle  \zeta _j^-\right),
\end{eqnarray}
where the expansion coefficients are of the form:
\begin{equation}\label{eq:3.46}
\begin{split}
\zeta_{s}^\pm(\lambda) &= -i\int_{-\infty}^\infty dx
\langle\hat\chi^{[s]\pm}(x,\lambda)|J|z(x)\rangle, \qquad
\zeta_{j}^\pm(\lambda)= -i\int_{-\infty}^\infty dx
\left\langle {\bf m}_j^\pm \right| J \left| z(x) \right\rangle .
\end{split}
\end{equation}

\begin{remark}
\label{rem:3.1}
If $q(x)\simeq 0$ then $\chi^+(x,\lambda)\simeq \chi^-(x,\lambda)
\simeq \exp(-i\lambda Jx)$ the set ${\cal  Z} $ is empty and
(\ref{eq:3.45}) goes into the usual Fourier transform for the space
$\bbbc^{n}$.
\end{remark}

\begin{remark}\label{rem:3.5}
Here we used also the fact that all eigenvalues of $J $ are non-vanishing.
In the case when one (or several) of them vanishes we can prove
completeness of the eigenfunctions only in a certain subspace of $\bbbc^n
$.
\end{remark}

The resolvent for the CBCs is defined quite analogously:
\begin{eqnarray}\label{eq:Re.1}
&& R(x,y,\lambda ) = R_\nu (x,y,\lambda ), \qquad \lambda \in \Omega _\nu
, \nonumber\\
&& R_\nu (x,y,\lambda ) = i \chi _\nu (x,\lambda ) \Theta ^\nu (x-y)
\hat{\chi} _\nu (x,\lambda ) , \nonumber\\
&& \Theta ^\nu (z) = \theta (-z) \Pi _0^\nu  - \theta (z) (\openone -\Pi
_0^\nu ), \qquad \Pi _0^\nu = \sum_{s \lenu k_{0,\nu }} E_{ss},
\end{eqnarray}
where $\chi _\nu (x,\lambda ) =\xi _\nu (x,\lambda ) e^{i\lambda Jx} $ and
$k_{0,\nu } $ is the number of positive eigenvalues of $\im (\lambda J) $
in the $\nu $-th ordering.

The following theorem is a specific case of one in \cite{GeYa}.
\begin{theorem}\label{th:GeYa}
Let $q(x) $ satisfy the conditions (C.1) and (C.2) and let ${\cal  Z}=
\cup_{p =1}^{n}\left({\cal  Z}_{2p-1}\cup {\cal  Z}_{2p}\right)  $ where
\begin{eqnarray}\label{eq:Re.3}
{\cal  Z}_{2p-1} \equiv \left\{ \lambda _j^+\omega ^{p-1} \in \Omega
_{2p-1}, \qquad j=1,\dots, N\right\}, \nonumber\\
{\cal  Z}_{2p} \equiv \left\{ \lambda _j^-\omega ^{p} \in \Omega
_{2p}, \qquad j=1,\dots, N\right\},
\end{eqnarray}
are the sets of zeroes and poles of the minors $m_{\nu ,k}(\lambda ) $ in
the sectors $\Omega _\nu  $. Then
\begin{enumerate}

\item $R_\nu (x,y,\lambda ) $ is an analytic function of $\lambda  $ for
$\lambda \in\Omega _\nu  $ having pole singularities at ${\cal  Z}_\nu  $;

\item $R_\nu (x,y,\lambda ) $ is a kernel of a bounded integral operator
for $\lambda \in \Omega _\nu  $;

\item For $\lambda \in l_\nu \cup l_{\nu +1} $ $R_\nu (x,y,\lambda ) $ is
an uniformly bounded function which is a kernel of an unbounded integral
operator;

\item $R_\nu (x,y,\lambda ) $ satisfies the equation:
\begin{equation}\label{eq:Re.5}
L(\lambda ) R_\nu (x,y,\lambda ) = \openone \delta (x-y).
\end{equation}
\end{enumerate}
\end{theorem}

The next natural step is to establish the structure of the singularities
of $R_\nu (x,y,\lambda ) $ at the points of ${\cal  Z} $. This is done
quite analogously by using the dressing factor (\ref{eq:26.2-n}). Note
that in these matters the symmetry complicates the calculations.

One of the effects of the $\bbbz_n $-symmetry is that the sets ${\cal
Z}_\nu  $ are determined uniquely by ${\cal  Z}_1 $ and ${\cal  Z}_{0} $:
\begin{equation}\label{eq:Re.6}
{\cal  Z}_{1} =\left\{ \lambda _j^+ \in \Omega _{1}, \quad j=1, \dots, N
\right\}, \qquad {\cal  Z}_{0} =\left\{ \lambda _j^- \in \Omega _{2n},
\quad j=1, \dots, N \right\}.
\end{equation}
The residue of $R_\nu (x,y,\lambda ) $ at the point $\lambda =\lambda
_j^{\nu } $ can be cast into the form:
\begin{eqnarray}\label{eq:Re.9}
\res_{\lambda =\lambda _j^+} R_{1} (x,y,\lambda ) = - 2i\im \lambda
_j^+ |{\bf n}_j^+ (x)\rangle \langle {\bf m}_j^+ (x) |, \nonumber\\
\res_{\lambda =\lambda _j^-} R_{2n} (x,y,\lambda ) = 2i\im \lambda
_j^- |{\bf n}_j^- (x)\rangle \langle {\bf m}_j^- (x) |,
\end{eqnarray}
where $|{\bf n}_j^\pm (x)\rangle  $ and $\langle {\bf m}_j^\pm (x)| $ are
properly normalized eigenvectors of the Lax operator corresponding to the
eigenvalues $\lambda _j^\pm \in \Omega _{\pm 1} $. The residues in the
other sectors $\Omega _\nu  $ with $\nu \neq 0, 1 (\mod 2n) $ are
evaluated from (\ref{eq:Re.9}) by employing eq. (\ref{eq:Z_n-cons}). Here
we also have the analog of the condition (\ref{eq:P_0-n}).

The derivation of the completeness relation of the eigenfunctions for CBCs
with $\bbbz_n $-reduction follows the same lines but needs some
modifications.  Instead of ${\cal J}(x,y) $ (\ref{eq:3.38}) we should
consider
\begin{equation}\label{eq:Re.10}
\tilde{{\cal  J}}(x,y) = \sum_{\nu =1}^{2n} {(-1)^{\nu -1}  \over 2\pi i }
\oint_{\gamma _\nu }^{} d\lambda R_\nu (x,y,\lambda ),
\end{equation}
where the contours $\gamma _\nu  $ are defined by:
\begin{equation}\label{eq:Re.11}
\gamma _{2\nu -1} = l_{2\nu -1}\cup \gamma _{2\nu -1}^{\infty } \cup
\bar{l}_{2\nu }, \qquad
\gamma _{2\nu } = \bar{l}_{2\nu }\cup l_{2\nu +1}^{\infty } \cup
\bar{\gamma }_{2\nu }.
\end{equation}
Here $l_\nu  $ are the rays (\ref{eq:l-nu}) oriented from $0 $ to $\infty
$; $\gamma _\nu ^{\infty } $ is the `infinite' arc $R_0e^{i\varphi _0} $
with $R_0\gg 1 $ and $\pi (\nu -1)/n \leq \varphi _0 \leq  \pi\nu /n $;
by overbar we denote the same contour with opposite orientation. Thus all
the contours $\gamma _{2\nu -1} $ (resp. $\gamma _{2\nu } $) are positively
(resp.  negatively) oriented.

Now we apply again the contour integration method and get two answers for
$\tilde{{\cal  J}}(x,y) $. The first, according to Cauchy residue theorem
is:
\begin{equation}\label{eq:Re.12}
\tilde{{\cal  J}}(x,y) = \sum_{p =1}^{2n} \sum_{j=1}^{N} \left(
\res_{\lambda =\lambda _j^+ \omega^p } R_{2p+1}(x,y,\lambda ) +
\res_{\lambda =\lambda _j^- \omega^p } R_{2p}(x,y,\lambda ) \right).
\end{equation}
Integration along the contours taking into account that
$\lim_{\lambda \to\infty }\chi ^\nu (x,\lambda ) =\openone $ gives:
\begin{equation}\label{eq:Re.13}
\tilde{{\cal  J}}(x,y) = \sum_{\nu =1}^{2n} {(-1)^{\nu -1}  \over 2\pi i }
\int_{l_\nu } dx \left( R_{\nu }(x,y,\lambda ) - R_{\nu -1}(x,y,\lambda )
\right) + J^{-1} \delta (x-y).
\end{equation}
The completeness relation follows after equating both expressions and
taking into account that (compare with (\ref{eq:Re.9}) and
(\ref{eq:Z_n-cons})):
\begin{eqnarray}\label{eq:Re.14}
\res_{\lambda =\lambda _j^+\omega ^p} R_{2p+ 1} (x,y,\lambda ) &=& - 2i\im
\lambda _j^+ |{\bf n}_j^{(2p+1)} (x)\rangle \langle {\bf m}_j^{(2p+1)} (x)
|, \nonumber\\
\res_{\lambda =\lambda _j^-\omega ^p} R_{2p} (x,y,\lambda ) &=& 2i\im
\lambda _j^+ |{\bf n}_j^{(2p)} (x)\rangle \langle {\bf m}_j^{(2p)} (x)|,
\end{eqnarray}
where $|{\bf n}_j^{(2p)} (x)\rangle$,  $|{\bf n}_j^{(2p+1)} (x)\rangle$
(resp. $ \langle {\bf m}_j^{(2p)} (x)| $ , $ \langle {\bf m}_j^{(2p+1)} (x)|
$) are properly normalized discrete eigenfunctions of the CBCs
(\ref{eq:2.2}) (resp. of the adjoint CBCs (\ref{eq:N16.2})) corresponding
to the discrete eigenvalues $\lambda _j^-\omega ^{2p} $ and $\lambda
_j^+\omega ^{2p} $. For the lack of space we can not provide all the
details of the calculations. The final result is similar to the one for
GZSs. Namely, any vector-function $|z(x)\rangle \in \bbbc^n $ can be
expanded over the eigenfunctions of the CBCs as follows:
\begin{eqnarray}\label{eq:Re.15}
|z(x)\rangle  = \\
\sum_{\nu =1}^{2n} {(-1)^{\nu -1}  \over 2\pi }
\int_{l_\nu }^{} d\lambda \left\{ \sum_{s\lnu k_{0,\nu }} \zeta _{\nu
,s}^{+}(\lambda ) | \chi ^{\nu ,[s]}(x,\lambda ) \rangle -
\sum_{s\gnu k_{0,\nu }} \zeta _{\nu ,s}^{-}(\lambda ) | \chi ^{\nu -1,
,[s]}(x,\lambda ) \rangle \right\} \nonumber\\
+ \sum_{j=1}^{N} \sum_{\nu =1}^{2n} \im \lambda _j^+ \left[
\zeta _{\nu ,j}^{+} | {\bf n}_j(x)^{\nu ,+} \rangle -
\zeta _{\nu ,j}^{-} | {\bf n}_j(x)^{\nu ,-} \rangle \right],
\nonumber
\end{eqnarray}
where the expansion coefficients are given by:
\begin{eqnarray}\label{eq:Re.16}
\zeta _{\nu ,s}^{+} (\lambda ) = -i \int_{-\infty }^{\infty } dx
\langle \hat{\chi }^{\nu ,[s]} (x,\lambda ) | J | z(x)\rangle ,
\nonumber\\
\zeta _{\nu ,s}^{-} (\lambda ) = -i \int_{-\infty }^{\infty } dx
\langle \hat{\chi }^{\nu-1 ,[s]} (x,\lambda ) | J | z(x)\rangle , \\
\zeta _{\nu ,j}^{+} = -i \int_{-\infty }^{\infty } dx \langle {\bf
m}_{j}^{\nu } (x)|J| z(x) \rangle , \qquad
\zeta _{\nu ,j}^{-} = -i \int_{-\infty }^{\infty } dx \langle {\bf
m}_{j}^{\nu -1 } (x)|J| z(x) \rangle . \nonumber
\end{eqnarray}

The completeness relations derived for GZSs and CBCs above can be
viewed as the spectral decompositions for the generically
non-self-adjoint operators $L(\lambda ) $.

\begin{remark}
The special case of a CBCs with $\bbbz_n$-symmetry is equivalent
to $n$-th order scalar differential operator \cite{DrinSok}.
Indeed, one can easily check that the system $L$ (\ref{eq:2.2}),
(\ref{eq:4.2}) can be written down as:
\begin{equation}\label{eq:L-chi}
L\chi \equiv i\left[ {d  \over dx } + \sum_{k=1}^{n}\psi _k(x)
K_0^k + i\lambda c_0\omega ^{-1/2} \sum_{k=1}^{n} \omega
^kE_{kk}\right] \chi (x,\lambda )=0.
\end{equation}
After similarity transformation with $u_0=\sum_{s,j=1}^{n} \omega
^{sj}E_{sj} $ goes into:
\begin{eqnarray}\label{eq:Lt-chi}
\tilde{L}\tilde{\chi} \equiv {1  \over i } u_0^{-1}Lu_0 \tilde{\chi}
\equiv \left[ {d  \over dx } + \sum_{s=1}^{n}\phi _s(x) E_{ss} -
\tilde{\lambda} \sum_{s=1}^{n} E_{s,s+1}\right] \tilde{\chi} (x,\lambda
)=0, \nonumber\\
\phi _s (x) = \sum_{k=1}^{n}\psi _k(x) \omega ^{ks} , \qquad
\tilde{\lambda } = i\lambda c_0\omega ^{-1/2},
\end{eqnarray}
and can be rewritten as the scalar operator
\begin{equation}\label{eq:rem1}
  L^{(n)}\chi_1 \equiv d_n d_{n-1} \cdots d_2 d_1
  \chi_1(x,\lambda) )  =\tilde{\lambda}^n   \chi_1(x,\lambda) ),
\end{equation}
where $d_k  X(x,\lambda)=  dX/dx + \phi_k (x) X(x,\lambda)$. If
$\phi_k (x)$ are real functions (additional $\bbbz_2$-reduction of
the type (\ref{eq:Z-N}) ensures this) then $ L^{(n)}$ is a
self-adjoint operator.

\end{remark}

\begin{remark}
The author is aware that these type of derivations need additional
arguments to be made rigorous. One of the real difficulties is to
find explicit conditions on the potential $q(x) $ that are
equivalent to the condition (C.2) or equivalently, to the
conditions that $m_k^\pm(\lambda ) $ have only finite number of
simple zeroes. Nevertheless there are situations (e.g., the
reflectionless potentials) when all these conditions are fulfilled
and all eigenfunctions of $L(\lambda ) $ can be explicitly
calculated. Another advantage of this approach is the possibility
to apply it to Lax operators with more general dependence on
$\lambda  $, e.g., quadratic or polynomial in $\lambda $.
\end{remark}

\section*{The `diagonal' of the resolvent}\label{sec:R(x,x)}

By the diagonal of the resolvent one usually means $R(x,y,\lambda
) $ evaluated at $y=x $. However the definition (\ref{eq:R2.1}) is
not continuous for $y=x $ and needs regularization. The simplest
possibility is to consider as the diagonal of the resolvent:
\[
R(x,\lambda ) = {1 \over 2 } \left( R(x+0,x,\lambda ) + R(x,x+0,\lambda
)\right).
\]
In fact we will consider as the a somewhat more general expression:
\begin{equation}\label{eq:R-P.1}
R_P(x,\lambda ) = i \chi ^{\pm}(x,\lambda ) P \hat{\chi }^\pm (x,\lambda
),
\end{equation}
where $P $ is a constant diagonal matrix. Obviously $R_P(x,\lambda ) $
satisfies
\begin{equation}\label{eq:R_p.2}
i {dR_P(x,\lambda )  \over dx } + \left[ q(x) - \lambda J, R_P(x,\lambda )
\right] \equiv [L(\lambda ), R_P(x,\lambda )] =0.
\end{equation}
Thus $R_P(x,\lambda ) $ belongs to the kernel of the operator
$[L(\lambda ),\cdot ] $. Due to the fact that $\chi ^\pm(x,\lambda
) $ is the FAS and satisfies a RHP with canonical normalization we
find:
\begin{equation}\label{eq:R-P.3}
R_P(x,\lambda ) = iP + \sum_{k=1}^{\infty } R_{P}^{(k)} (x) \lambda ^{-k}.
\end{equation}
The coefficients $ R_{P}^{(k)} (x) $ can be expressed through $q(x) $
using the recursion relations generalizing the ones of AKNS
\cite{AKNS,Ge2,Konop,GeKu}.  These relations are solved by the recursion
operators $\Lambda _\pm $ which have the form:
\begin{equation}\label{eq:7.7L}
\Lambda_\pm X= \ad_J^{-1} \left( i {dX \over dx } + P_0([q_x),X(x)] +
i\left[ q(x), \int_{\pm \infty }^{x} dy [q(y),X(y)]\right] \right),
\end{equation}
where $P_0 $ is the projector onto the off-diagonal part of the matrix
$P_0X=X_0^{\rm f} $, the matrix $X(x) $ in (\ref{eq:7.7L}) satisfies
$X\equiv P_0X $ and
\[ (\ad_{J}^{-1}X_0^{\rm f})_{ij} = {(X_0^{\rm f})_{ij} \over a_i-a_j}.
\]
The coefficients $R_{P}^{(k)} (x)  $ can be expressed in compact form
through $\Lambda _\pm $ as follows:
\begin{eqnarray}\label{eq:7.8}
R_P^{k+1}{}^{\rm f} &=& \Lambda_\pm R_P^{k}{}^{\rm f} =-\Lambda_\pm ^{k}
\ad_J^{-1} [iP,q(x,t)], \\
\label{eq:7.8'}
R_P^{(k)}{}^{\rm d} &=& i \int_{\pm\infty }^{x} dy (\openone -P_0) \left(
[q(y,t), R_P^{(k)}{}^{\rm f}]\right) + \lim_{x\to\pm\infty }
R_P^{(k)}{}^{\rm d}(x,t).
\end{eqnarray}

Quite naturally these coefficients, or rather the diagonal of the
resolvent generates \cite{GelDick,Dick,Ge2}:

-- the class of NLEE. Given the dispersion law, e.g., $f(\lambda )=\lambda
^N P $ of the NLEE we can write down the equation itself by:
\begin{equation}\label{eq:7.9}
-i{dq  \over dt } + i {\left(dR_P^{(N)}\right)^{\rm f} \over dx } +
P_0([q(x,t),R_P^{(N)}(x,t)] =0.
\end{equation}

-- the corresponding Lax representations, or in other words, the $M
$-operators for each of these NLEE as follows:
\begin{equation}\label{eq:R-P.4}
V_P^{(N)}(x,\lambda ) = \sum_{k=0}^{N} R_P^{(k)} (x) \lambda ^{N-k}.
\end{equation}

-- the integrals of motion of the corresponding NLEE. This follows from

\begin{theorem}[\cite{Ge2}]\label{th:LMP6}
The quantities
\begin{equation}\label{eq:R-P.6}
R_{\Pi ^{(k)}}^{\pm} (x,\lambda ) = i \chi ^\pm(x,\lambda ) \Pi ^{(k)}
\hat{\chi} ^\pm(x,\lambda ), \qquad \Pi ^{(k)}= \sum_{s=1}^{k} E_{ss} - {
k \over n } \openone ,
\end{equation}
satisfy the relations
\begin{equation}\label{eq:R-P.7}
\int_{-\infty }^{\infty } dx \tr \left( R_{\Pi ^{(k)}}^{\pm} (x,\lambda )
J - i\Pi ^{(k)}J \right) = - {d  \over d\lambda  } {\cal  D}_k(\lambda ),
\end{equation}
where ${\cal  D}_k^\pm(\lambda ) $ is defined by (\ref{eq:N23.3}).

\end{theorem}

Combined with the (\ref{eq:7.7L}) we can deduce that the diagonal of the
resolvent and the recursion operator
\begin{equation}\label{eq:R-P-L}
(\Lambda _\pm - \lambda ) R_P^\pm (x,\lambda ) = i [P,\ad_J^{-1} q(x)],
\end{equation}
directly reproduce the generating functionals of the conserved quantities.

The termin `squared' solutions and recursion operator do not reflect
properly the algebraic properties of these objects.  The recursion
operators $\Lambda _\pm $ can be understood as the Lax operator $L(\lambda
) $ in the adjoint representation. One of the definitions of the adjoint
representation means that we should replace each element $U(x,\lambda )\in
\fr{g} $ by $\ad_{U(x,\lambda )}\cdot = [U(x,\lambda ),\cdot] $. Therefore
due to (\ref{eq:R_p.2}) we can view the diagonal of the resolvent
$R_P^\pm(x,\lambda ) $ as the eigenfunction of $L(\lambda ) $ in the
adjoint representation. It remains to project out the kernel of $\ad_J $
in order to derive $\Lambda _\pm $ from $L(\lambda ) $.

The `squared' solutions are eigenfunctions of $\Lambda _\pm $ and belong
to a linear space, which is the co-adjoint orbit of $\hat{\fr{g}}^*_+ $
determined by $J $.  The gauge covariant way to introduce them involves
the FAS of $L(\lambda ) $ and is:
\begin{equation}\label{eq:R-P.8}
\qquad e_{ij}^{\pm} (x,\lambda ) = P_0 \left(\chi ^\pm(x,\lambda )
E_{ij}\hat{\chi }^\pm(x,\lambda ) \right), \qquad
h_{j}^{\pm} (x,\lambda ) = P_0 \left(\chi ^\pm(x,\lambda ) H_j
\hat{\chi }^\pm(x,\lambda ) \right),
\end{equation}
where $\chi ^\pm(x,\lambda ) $ are the FAS of $L(\lambda ) $ GZSs. The
similarity transformation by $\chi ^\pm(x,\lambda ) $ is the adjoint
action of the group $\fr{G} $ on the algebra $\fr{g} $; therefore
$e_\alpha ^\pm(x,\lambda ) $ and $h_j^\pm(x,\lambda ) $ are elements again
of $\fr{g} $. The projection $\Pi _0 =\ad_J^{-1}\ad_J $ is a natural
linear operator on $\fr{g} $. Besides the `squared'
solutions are analytic functions of $\lambda  $ having both poles and
zeroes at $\lambda _j^\pm $.

More detailed analysis  based on the Wronskian relations reveals
several other important aspects \cite{KauNew,Ge,GeYa} of the
`squared' solutions of GZSs.  First,  the sets
\[ \{ e_{ij}^{+}(x,\lambda ),  e_{ji}^{-}(x,\lambda)\}, \; e_{ij;k}^{+}(x),
e_{ji;k}^{-}(x), \dot{e}_{ij;k}^{+}(x), \dot{e}_{ji;k}^{-}(x),
\qquad i<j , k=1,\dots N \}
\]
and
\[ \{ e_{ji}^{+}(x,\lambda ), e_{ij}^{-}(x,\lambda)\},\; e_{ji;k}^{+}(x),
e_{ij;k}^{-}(x), \dot{e}_{ji;k}^{+}(x), \dot{e}_{ij;k}^{-}(x),
\qquad i<j , k=1,\dots N \}
\]
form complete sets of functions on ${\cal  M} $ that realize the
mapping ${\cal M}\leftrightarrow  {\cal T} $. Here by $
e_{ji;k}^{\pm}(x)  $ and $\dot{e}_{ji;k}^{\pm}(x)  $ we have
denoted:
\[
e_{ij;k}^{\pm}(x)  = e_{ij}^{\pm} (x,\lambda _k^\pm), \qquad
\dot{e}_{ij;k}^{\pm}(x) =\left. {d e_{ij}^{\pm} (x,\lambda )  \over
d\lambda  } \right|_{\lambda =\lambda _k^\pm}.
\]

Second, it is possible to expand the potential $[P, \ad_J^{-1}
q(x,t)] $ and its variation $\ad_J^{-1}\delta q(x) $ over each of the
complete sets shown above. The corresponding expansion coefficients are
expressed through ${\cal T} $ and their variations. These facts constitute
the grounds on which one can show that the ISM can be understood as a
generalized Fourier transform. The important difference as compare to the
standard Fourier transform is in the fact that the operator $L $
(as well as the operators $\Lambda _\pm $) allows for discrete
eigenvalues. Therefore the completeness relations involve both
integrals along the continuous spectrum and sum over the discrete
eigenvalues. In the usual Fourier transform the discrete
eigenvalues are absent.

\section*{Hamiltonian properties of the NLEE}\label{sec:3.2}

Here we briefly formulate the Hamiltonian properties of the NLEE
paying more attention to its algebraic structure. This has been
widely studied problem, see
\cite{Adler,DrinSok,KuRe,FlaNewRat,GelDick,Dick,FaTa,ZMNP,Ge2,Ge}
and the numerous references therein.

In doing so we follow mainly the ideas of \cite{KuRe} with  a
natural generalization from $sl(2) $ to $sl(n) $-algebras. The
main idea in these papers is the possibility to write down the Lax
equation (\ref{eq:lax}) in explicitly Hamiltonian form as the
co-adjoint action of $\tilde{\fr{g}} $ on its dual
$\tilde{\fr{g}}^* $. Obviously any non-trivial grading in $\fr{g}
$ (resp. $\tilde{\fr{g}} $, $\hat{\fr{g}} $) will reflect into a
corresponding grading of the dual algebra $\fr{g}^* $ (resp.
$\tilde{\fr{g}}^* $, $\hat{\fr{g}}^* $).

Below we will need also the Cartan-Weyl basis of $sl(n) $. Choosing for
definiteness the typical $n\times n $ representation we fix it up by:
\begin{equation}\label{eq:CW-sln}
\fr{h} \equiv \mbox{l.c.} \left\{ H_i = E_{ii}-E_{i+1,i+1}, \quad
i=1,\dots, n-1 \right\} , \qquad \left\{ E_{ij}, \quad i\neq j\right\}.
\end{equation}
As invariant bilinear form we can use $\langle X,Y\rangle =\tr(XY) $.
Then the commutation relations can be written in the form:
\begin{eqnarray}\label{eq:CW-cr}
[H_i , E_{jk}] &=& (e_i-e_{i+1}, e_j-e_k) E_{jk}, \qquad j\neq k,
\nonumber\\
\left[ E_{jk} , E_{kl}\right] &=& E_{jl}, \qquad [E_{jk} , E_{lj}] =-
E_{lk}, \qquad l\neq j, \\
\left[E_{jk} , E_{kj}\right] &=& \sum_{s=j}^{k-1} H_{s}, \qquad j<k.
\nonumber
\end{eqnarray}
By $e_k $ above we mean an orthonormal basis in the $n
$-dimensional Euclidean space with a standard scalar product:
$(e_j,e_k)=\delta _{jk} $. Those, who are familiar with Lie algebras will
recognize $e_i-e_{i+1} $ as the simple roots of $sl(n) $ and the set of
$e_j-e_k $, $j\neq k $ as the root system of $sl(n) $.

If $C=\openone  $ (i.e. with the trivial grading) each of the matrices
$U_k(x) $ in (\ref{eq:1.18}) is of generic form:
\begin{equation}\label{eq:u-gen}
U_k(x) = \sum_{j=1}^{n-1}u_{j}^{(k)} H_j + \sum_{j\neq p} u_{jp}^{(k)}
E_{jp}.
\end{equation}
The coefficients $u_{j}^{(k)}(x) $, $u_{jp}^{(k)}(x) $ can be
viewed as linear functionals on $u_k(x) $ and thus they belong to
$\fr{g}^* $. Using the bilinear form (\ref{eq:Kill}) they can be
interpreted as linear functionals on $\hat{\fr{g}} $ and thus as
elements also of $\hat{\fr{g}}^* $. The algebraic structure on
$\tilde{\fr{g}}^* $ can be introduced in analogy with the
commutation relations (\ref{eq:CW-cr}), namely:
\begin{eqnarray}\label{eq:PB}
\left\{ u_{i}^{(s)}(x), u_{j,j+k}^{(m)}(y) \right\}_{p} &=& (e_i-e_{i+1},
e_j-e_k) u_{j,j+k}^{(s+m-p)}(x) \delta (x-y), \nonumber\\
\qquad \left\{ u_{i,i+k}^{(s)}(x), u_{i+k,j}^{(m)}(y) \right\}_{p}
&=&
u_{i,j}^{(s+m-p)}(x) \delta (x-y), \\
\left\{ u_{i,i+k}^{(s)}(x), u_{j+k,i}^{(m)}(y) \right\}_{p} &=&
-u_{j+k,i+k}^{(s+m-p)}(x) \delta (x-y), \nonumber\\
\left\{ u_{i,i+k}^{(s)}(x), u_{i+k,i}^{(m)}(y) \right\}_{p} &=&
\sum_{l=i}^{i+k-1} u_{l}^{(s+m-p)}(x) \delta (x-y) + i \delta
_{s+m,p}\delta ' (x-y). \nonumber
\end{eqnarray}
The derivation of these relations follows \cite{KuRe} in a rather
straightforward manner; though a bit tedious, it can be generalized also
to any simple Lie algebra.

Note that if $p=-1 $ then the term with $\delta '(x-y) $ disappears and
the Poisson brackets (\ref{eq:PB}) become ultralocal. Then we can rewrite
them in a compact form using the classical $r $-matrix \cite{FaTa}:
\begin{eqnarray}\label{eq:r-mat.1}
\qquad \left\{ U(x,\lambda )\otimescomma U(y,\mu ) \right\}_{-1} =
[ r(\lambda -\mu ) , U(x,\lambda ) \otimes \openone  + \openone
\otimes U(x,\mu )] \delta
(x-y), \\
r(\lambda -\mu ) = {\Pi_0  \over \lambda -\mu  }, \qquad  \Pi_0 =
\sum_{i,j=1}^{n} E_{ij}\otimes E_{ji}.
 \end{eqnarray}
The left hand side of (\ref{eq:r-mat.1}) has the structure of the usual
tensor product of $n\times n $ matrices, but instead of taking the product
one should rather take the Poisson bracket between the corresponding matrix
elements of $U(x,\lambda ) $ and $U(y,\mu ) $.

The relations (\ref{eq:r-mat.1}) are local in the sense that for
the evaluation of the left hand side of (\ref{eq:r-mat.1}) we need
to use only the Poisson brackets between the matrix elements of
$U(x,\lambda ) $ and do not need the boundary conditions on the
potentials. The effectiveness of the $r $-matrix, when it exists,
is in the possibility to evaluate the Poisson brackets between the
matrix elements of the scattering matrix $T(\lambda ) $. To do
this we need to `integrate' (\ref{eq:r-mat.1}) which needs to take
into account also the boundary conditions. For periodic boundary
conditions on $q(x) $ this  gives:
\begin{equation}\label{eq:TT-per}
\left\{ T(\lambda ) \otimescomma T(\mu ) \right\}_{-1} =
[r(\lambda -\mu ), T(\lambda ) \otimes T(\mu )].
\end{equation}
{}For vanishing boundary conditions on $q(x) $ and $J=J^* $ the
calculations need some additional considerations with the result
(see \cite{FaTa}):
\begin{eqnarray}\label{eq:TT-va}
\left\{ T(\lambda ) \otimescomma T(\mu ) \right\}_{-1} =
r_+(\lambda -\mu ) T(\lambda ) \otimes T(\mu ) - T(\lambda )
\otimes T(\mu ) r_-(\lambda -\mu
), \nonumber\\
r_\pm(\lambda -\mu ) = {1 \over \lambda -\mu  } \sum_{j=1}^{n} E_{jj}
\otimes E_{jj} \mp i\pi \delta (\lambda -\mu ) \sum_{i\neq j=1}^{n}
E_{ij} \otimes E_{ji}.
\end{eqnarray}
{}From both relations (\ref{eq:TT-per}) and (\ref{eq:TT-va}) there
follows that the principal minors $m_k^\pm(\lambda ) $ commute
with respect to the Poisson brackets (\ref{eq:PB}) \cite{Ge},
i.e.:
\begin{equation}\label{eq:pb-m}
\{ {\cal D}_k (\lambda ), {\cal D}_j(\mu )\}_{-1} =0.
\end{equation}
Since ${\cal D}_k(\lambda ) $ are the generating functionals of
integrals of motion ${\cal D}_k^{(s)}$ (see eq.
(\ref{eq:as-exp})), then eq. (\ref{eq:pb-m}) means that all these
integrals are in involution with respect to these Poisson
brackets.

The $\bbbz_n$-symmetry may modify substantially some of the above
results. Indeed, it can be viewed as a set of constraints on the
phase space ${\cal M}$ and on the generic Poisson brackets
(\ref{eq:PB}). Then one should evaluate the corresponding Dirac
brackets on the reduced phase space. However in the case of the
$\bbbz_n$-NLS equation (\ref{eq:3.3}) with Lax operator $L$ given
by (\ref{eq:2.2}), (\ref{eq:4.2}) somewhat surprisingly the
approach of \cite{KuRe} gives us directly the correct answer. If
we define $\psi_j(x,t)$ as linear functionals of $U(x,t,\lambda) =
q(x,t) -\lambda J $  by:
\begin{equation}\label{eq:psi-j}
  \psi_j(x,t) = {1\over n} \tr \left( U(x,t,\lambda) K^{n-j}
  \right) ,
\end{equation}
and make use of (\ref{eq:4.2}) then the set of Poisson brackets in
(\ref{eq:PB}) simplify to
\begin{equation}\label{eq:pb1}
\left\{  \psi_j(x,t) ,  \psi_k(y,t) \right\} = \delta_{k+j-n}
\delta' (x-y).
\end{equation}
Together with the Hamiltonian $H=\omega^2 M_{1,1}^{(2)}$
(\ref{eq:2.51''}) they provide the Hamiltonian formulation of
(\ref{eq:3.3}). Unfortunately this Poisson brackets are not
ultra-local and the corresponding Lax operator does not allow
classical $r$-matrix of the form (\ref{eq:r-mat.1}).

{}For the affine Toda chain (\ref{eq:3.2}) the simplest Poisson
brackets are provided by:
\begin{equation}\label{eq:pb-to}
\left\{ {d Q_j\over dx} ,  Q_k(y,t) \right\} = \delta_{kj} \delta
(x-y).
\end{equation}
The corresponding Lax operator (\ref{eq:To1}) unlike the previous
case allows classical $r$-matrix satisfying (\ref{eq:r-mat.1})
which however has more complicated dependence on  $\lambda -\mu$;
it is known as the trigonometric $r$-matrix \cite{PPK}.

Another special property of the $\bbbz_n$-symmetric CBCs concerns
the existence of the so-called symplectic basis \cite{GeKh}. The
elements of these bases are special linear combinations of the
`squared solutions' (\ref{eq:R-P.8}) which are also complete in
${\cal M}$ and which are such that the expansion coefficients of
$\delta q(x,t)$ over it produce the variations of the action-angle
variables of the corresponding set of NLEE. In \cite{GeKh} this
basis was worked out for the Zakharov-Shabat system related to the
$sl(2)$ algebra. For GZSs related to algebras of higher rank such
basis is yet unknown although it must exist since the
action-angle variables for them are known \cite{Man,BeSat}.

{}For the $\bbbz_n$-symmetric CBCs the construction of the
symplectic basis is very much like the one in \cite{GeKh}  due to
the fact that the subalgebras ${\cal g}_\nu$ related to each of
the rays $l_\nu$ are direct sums of $sl(2)$ subalgebras. It is a
complete set of functions on the phase space of the corresponding
$\bbbz_n$-symmetric NLEE (\ref{eq:3.1}) and (\ref{eq:3.2}).
Skipping the details we just give the explicit expressions for the
set ${\cal A}$ of action-angle variables of the $\bbbz_n$-NLS
equation in terms of the scattering data of its Lax operator.
Obviously ${\cal A}$ will consists of two sets of functions ${\cal
A}= {\cal A}_0\cup {\cal A}_1$ each set defined on the ray $l_0$
and $l_1$ respectively:
\begin{eqnarray}\label{eq:A-0-1}
{\cal A}_0 &\equiv & \left\{ \pi_{ij}(\lambda),
\kappa_{ij}(\lambda), \qquad \lambda \in l_0, \qquad i+j=2 (\mod
n) \right\}, \nonumber \\
{\cal A}_1 &\equiv & \left\{ \pi_{ij}(\lambda),
\kappa_{ij}(\lambda), \qquad \lambda \in l_1, \qquad i+j=1 (\mod
n) \right\}, \nonumber
\end{eqnarray}
where
\begin{equation}\label{eq:pi-kap}
\pi_{ij} (\lambda) = -{1\over \pi} \ln \left( 1+\rho_{ij}^+
\rho_{ij}^-\right), \qquad \kappa_{ij} (\lambda) = -{i\over 2}
{b_{ij}^+(\lambda)\over b_{ij}^-(\lambda)}, \qquad
\rho_{ij}^+(\lambda) = {b_{ij}^+(\lambda)\over a_{ij}^+(\lambda)},
\end{equation}
and the coefficients $a_{ij}^+(\lambda)$, $b_{ij}^+(\lambda)$ were
introduced in (\ref{eq:T-km}).

Quite analogous are  the expressions for the action-angle
variables for the two-dimensional affine Toda chain provided we
use the scattering data of the Lax operator (\ref{eq:To2}).

\specialsection{Conclusion}\label{sec:4}

The restricted space did not allow us to give more details or
explanations on these and related problems. We only mention some of them
below.

One such important to our mind result is the interpretation of the
ISM as a generalized Fourier transform.  In its derivation for the GZSs
and CBCs \cite{GeKu,Ge,GeYa} both algebraic methods and analytic ones were
used. As a result the pair-wise equivalence of the symplectic structures
in the hierarchy becomes obvious.

The approach based on the Kac-Moody algebras is a natural basis for the
Hamiltonian hierarchies. If one can derive a bi-Hamiltonian formulation of
a given NLEE then there is a whole hierarchy of them related by a
recursion operator $\Lambda  $. Here we mention the paper \cite{FokAnd}
where the operator $\Lambda  $ was derived as the `ratio' of two such
Hamiltonian structures for the $N $-wave equations. The result, of course
coincides with the natural expression for $\Lambda $ obtained with the
AKNS recursion method and whose spectral theory was constructed by other
means in \cite{GeKu,Ge2}.

The method based on the diagonal of the resolvent of the Lax operator
started by Gel'fand and Dickey \cite{GelDick,Dick} can be viewed also as a
formal algebraic one. The authors studied by algebraic means the ring of
operators, commuting with $L $. They expressed most of the quantities,
including the diagonal of the resolvent of $L $, as series over fractional
powers of $L $ and did not investigate the existence and convergence
of these series.  Once identified with the expression (\ref{eq:R-P.1}) in
terms of the FAS these problems find their natural and positive solution.

Besides the classical $r $-matrix corresponding to the ultralocal Poisson
brackets there exist also dynamical $r $-matrices depending on the fields
$q_{ij}(x) $ in the NLEE. One of the problems, that is still not solved is
to find the interrelation between the dynamical $r $-matrices, $r $ and
the recursion operator $\Lambda  $.

{}Finally, we should mention that both approaches have been
further generalized.  For example, the analytic approach was
generalized from a local RHP to a nonlocal RHP and to $\partial
$-bar problem (also local and nonlocal), see
\cite{AblFokAn,ZaMa133,Konop}. This allowed to treat NLEE of
soliton type in $2+1 $ dimensions.

Another direction is to study Lax operators with more general
$\lambda  $-dependence such as polynomial, or rational
\cite{ZaMi}.

Obviously all results concerning spectral decompositions can be formulated
in a gauge covariant way thus allowing to treat also gauge equivalent NLEE
\cite{GeYa*85,GeYa*86,Ge}.

The algebraic approach was also generalized to use as a basis infinite
dimensional algebras such as Virasoro algebra, $W_{1+\infty } $ etc. which
lead to the important construction of the Japanese $\tau $-function and
its relation to the soliton theory, see \cite{JiMi,FlaNewRat}.

Thus we just outlined the beginning of all this and so it is time to stop.

\section*{Acknowledgements}\label{sec:ack}
The author wishes to thank Professor D. J. Kaup and Professor R.
Choudhurry for financial support making his participation in the
conference possible as well as for useful discussions. I am
grateful to Professors  G. Vilasi, S. de Filippo and M. Salerno
for useful discussions and warm hospitality at Salerno
University where this paper was finished.

\appendix

\section{Gauss decompositions}\label{sec:A1}

The Gauss decompositions mentioned above have natural group-theoretical
interpretation and can be generalized to any semi-simple Lie algebra. It
is well known that if given group element allows Gauss decompositions then
its factors are uniquely determined. Below we write down the explicit
expressions for the matrix elements of $T^\pm(\lambda ) $, $S^\pm(\lambda
) $, $D^\pm(\lambda ) $ through the matrix  elements of $T(\lambda ) $:
\begin{eqnarray}\label{eq:1.50a}
T^-_{pj}(\lambda ) &=& {1 \over m_{j}^+(\lambda ) } \left\{
\begin{array}{ccccc} 1, & 2, & \dots , & j-1, & p \\ 1, & 2, & \dots , &
j-1, & j \end{array}\right\}_{T(\lambda )}^{(j)}, \\
\hat{T}^-_{jp}(\lambda ) &=& {(-1)^{j+p}  \over m_{j-1}^+(\lambda )}
\left\{\begin{array}{cccccc} 1, & 2, & \dots ,&\check{p}, & \dots , & j \\
1, & 2, & \dots , & p, & \dots , & j-1 \end{array}\right\}_{T(\lambda
)}^{(j-1)}, \\
S^+_{pj}(\lambda ) &=& {(-1)^{p+j} \over m_{j-1}^+(\lambda ) } \left\{
\begin{array}{cccccc} 1, & 2, & \dots , & p, &\dots, & j-1 \\ 1, & 2, &
\dots , & \check{p},& \dots, & j \end{array}\right\}_{T(\lambda)}^{(j)},
\\
\hat{S}^+_{jp}(\lambda ) &=& {1  \over m_{j}^+(\lambda )}
\left\{\begin{array}{ccccc} 1, & 2, & \dots ,& j-1, & j \\
1, & 2, & \dots , & j-1, & p \end{array}\right\}_{T(\lambda )}^{(j-1)},\\
T^+_{pj}(\lambda ) &=& {1 \over m_{n-j+1}^-(\lambda ) } \left\{
\begin{array}{ccccc} p, & j+1, & \dots , & n-1, & n \\ j, & j+1, & \dots ,
& n-1, & n \end{array}\right\}_{T(\lambda )}^{(n-j+1)}, \\
\hat{T}^+_{jp}(\lambda ) &=& {(-1)^{p+j} \over m_{n-j}^-(\lambda ) }
\left\{ \begin{array}{cccccc} j, & j+1, &\dots, &\check{p}, & \dots , & n
\\ j+1, & j+2, & \dots ,&p, & \dots, & n \end{array}\right\}_{T(\lambda
)}^{(n-j)}, \\
S^-_{pj}(\lambda ) &=& {(-1)^{p+j} \over m_{n-j}^-(\lambda ) }
\left\{ \begin{array}{cccccc} j+1, & j+2,&\dots, &p, & \dots , & n \\
j, & j+1, & \dots ,&\check{p}, & \dots, & n \end{array}\right\}_{T(\lambda
)}^{(n-j)}, \\
\hat{S}^-_{jp}(\lambda ) &=& {1 \over m_{n-j+1}^-(\lambda ) }
\left\{ \begin{array}{ccccc} j, & j+1, & \dots ,&n-1, & n \\
p, & j+1, & \dots , & n-1, & n \end{array}\right\}_{T(\lambda)}^{(n-j+1)},
\end{eqnarray}
where
\begin{equation}\label{eq:10.7}
\left\{ \begin{array}{cccc} i_1, & i_2, & \dots ,&i_k, \\
j_1, & j_2, & \dots , & j_k \end{array}\right\}_{T(\lambda)}^{(k)} =
\det \left| \begin{array}{cccc} T_{i_ij_1} & T_{i_1j_2} & \dots T_{i_1j_k}
\\ T_{i_2j_1} & T_{i_2j_2} & \dots T_{i_2j_k} \\ \vdots & \vdots & \ddots
& \vdots \\ T_{i_kj_1} & T_{i_kj_2} & \dots T_{i_kj_k} \end{array} \right|
\end{equation}
is the minor of order $k $ of $T(\lambda ) $ formed by the rows $i_1 $, $
i_2 $, \dots , $i_k $ and the columns $j_1 $, $j_2 $, \dots, $j_k $; by $
\check{p} $ we mean that $p $ is missing.

{}From the formulae above we arrive to the following
\begin{corollary}\label{cor:1.1}
In order that the group element $T(\lambda ) \in SL(n,\bbbc) $
allows the first  (resp. the second)  Gauss decomposition
(\ref{eq:15.1}) is necessary and sufficient that all upper- (resp.
lower-) principle minors $m_k^+(\lambda ) $ (resp. $m_k^-(\lambda
) $) are not vanishing.
\end{corollary}

These formulae hold true also if we need to construct the Gauss
decomposition of an element of the orthogonal $SO(n) $ group. Here we
just note that if $T(\lambda )\in SO(n) $ then
\begin{equation}\label{eq:def-G}
S_0 (T(\lambda ))^T S_0^{-1} = T^{-1}(\lambda ),
\end{equation}
where
\begin{eqnarray}\label{eq:def-S0}
S_0 = \sum_{k=1}^{n_0} (-1)^{k+1} (E_{k,n+1-k} + E_{n+1-k,k}), \quad
\text{if}\quad n=2n_0, \\
S_0 = \sum_{k=1}^{n_0} (-1)^{k+1} (E_{k,n+1-k}+E_{n+1-k,k}) +
(-1)^{n_0}E_{n_0+1,n_0+1} , \quad \text{if}\quad n=2n_0+1.
\nonumber
\end{eqnarray}
One can check that if $T(\lambda ) $ satisfies (\ref{eq:def-G})
then each of the factors $T^\pm(\lambda ) $, $S^\pm(\lambda ) $
and $D^\pm(\lambda ) $ also satisfy (\ref{eq:def-G}) and thus
belong to the same group $\fr{G} $. In addition we have the
following interrelations between the principal minors of
$T(\lambda ) $:
\begin{eqnarray}\label{eq:m_k}
m_j^\pm(\lambda ) = m_{n-j}^\pm(\lambda ), \qquad \text{for} \quad SO(n),
\nonumber\\
m_j^\pm(\lambda ) = m_{n-j}^\pm(\lambda ), \qquad \text{for} \quad SP(n),
\end{eqnarray}

\section{Dispersion relations for  ${\cal D}_k(\lambda ) $ and ${\ln m}_{\nu,k}^+(\lambda ) $}\label{sec:A2}

Let us introduce the functions $f_k^\pm(\lambda ) $:
\begin{eqnarray*}
f_k^+(\lambda ) = {  m_k^+(\lambda )\over R_k (\lambda ) } , \qquad
f_{n-k}^-(\lambda ) = R_k(\lambda ) m_{n-k}^-(\lambda ), \qquad
R_k(\lambda )=\prod_{j=1}^{N} \left({\lambda -\lambda _j^+ \over \lambda
-\lambda _j^- } \right)^{b_{jk}} ,
\end{eqnarray*}
which like $m_k^\pm(\lambda ) $ are: i) analytic for $\lambda \in
\bbbc_\pm $; ii) satisfy $\lim_{\lambda \to\infty } f_k^\pm(\lambda ) =1
$. Besides, $f_k^\pm(\lambda ) $ have no zeroes in their regions of
analyticity and therefore the functions $\ln f_k(\lambda ) $ are analytic
for $\lambda \in \bbbc_\pm $ and tend to $0 $ for $\lambda \to \infty  $.
This allows one to apply the Plemelji-Sokhotzky formula with the result:
\begin{align}\label{eq:N23.3'}
\tilde{{\cal  D}}_k(\lambda ) &= {1 \over 2\pi i } \int_{-\infty }^{\infty
} { d\mu  \over \mu -\lambda  } \ln \left( f_k^+(\mu ) f_{n-k}^-(\mu )
\right), \\
\intertext{where}
\label{eq:N23.4'}
\tilde{{\cal  D}}_k(\lambda ) &= \left\{ \begin{array}{ll} \ln
f_k^+(\lambda ), \qquad & \lambda \in \bbbc_+ \\ -\ln f_{n-k}^-(\lambda ),
\qquad & \lambda \in \bbbc_- .\end{array} \right.
\end{align}
It remains to insert the above definitions of $f_k^\pm(\lambda ) $ into
(\ref{eq:N23.3'}) and (\ref{eq:N23.4'}) to  derive Eqs.  (\ref{eq:N23.3}),
(\ref{eq:N23.4}).

The dispersion relation  (\ref{eq:dis-rel}) is derived analogously
treating the integral
\begin{equation}\label{eq:tt-J}
 \tilde{\tilde{\cal J}}(\lambda) = \sum_{\nu=1}^{2n} {(-1)^{\nu-1}\over 2\pi
 i} \oint_{\gamma_i} {d\mu \over \mu - \lambda} \ln \left\{
 m_{\nu,k}^+(\mu) \prod_{\eta=1}^n \prod_{j=1}^N \left( \mu -
 \lambda_{j,k}^+ \over \mu - \lambda_{j,k}^- \right)
 ^{b_{k,j}^\eta} \right\} ,
\end{equation}
with $\lambda \in \Omega_\nu$ and the contours $\gamma_i $ as in
(\ref{eq:Re.11}).

\bibliographystyle{amsalpha}

\begin{thebibliography}{A}

\bibitem{AblFokAn}
M.~J.Ablowitz, A.~S.~Fokas, R.~Anderson.
\textit{The direct linearising transform and the Benjamin--Ono equation.}
Phys.\ Lett.\ {\bf 93A}, n.8, 375--378, 1983.

\bibitem{AKNS}
M.~J.~Ablowitz, D.~J.~Kaup, A.~C.~Newell, H.~Segur.
\textit{The inverse scattering transform -- Fourier analysis for nonlinear
   problems.}
Studies in Appl.\ Math.\ {\bf 53}, n.~4, 249--315, 1974.

\bibitem{Adler} M. Adler. \textit{On a trace functional for formal
pseudo-differential operators and the symplectic structure of the
Korteweg-de Vries equations.} Inv.  Math.  {\bf 50}, 219-248 (1979).

\bibitem{AG} N. I. Akhiezer, I. M. Glazman.  \textit{Theory of Linear operators in Hilbert
space.}  Translated from Russian, New York, F. Ungar (1961-1963).


\bibitem{BeCo1} R.~Beals, R.~R.~Coifman. \textit{ Scattering and inverse scattering for
first order systems.} Commun. Pure \& Appl. Math. {\bf 37}, 39
(1984).

\bibitem{BeCo2} R.~Beals, R.~R.~Coifman. \textit{Inverse scattering and evolution equations.}
Commun. Pure \& Appl. Math. {\bf 38}, 29 (1985).

\bibitem{BeSat} R.~Beals, D.~H.~Sattinger. \textit{On the complete integrability of completely
integrable systems.} Commun. Math. Phys. {\bf 138}, 409 (1991).

\bibitem{Caud} P.~J.~Caudrey. \textit{The inverse problem for the third order equation
$u_{xxx} + q(x)u_x + r(x)u = -i\zeta^3 u$.} Phys. Lett.~A {\bf 79A}, 264 (1980); \\
--- \textit{The inverse problem for a general $n\times n$ spectral
equation. } Physica D {\bf D6}, 56 (1982).

\bibitem{CalDeg} F. Calogero, A. Degasperis. \textit{ Spectral  transform
and  solitons. Vol.~I.\,} North Holland, Amsterdam, 1982.

\bibitem{Dick} L. A. Dickey. \textit{Soliton equations and Hamiltonian
systems.} Advanced series in Math. Phys., {\bf 12}, World Scientific,
(1991).

\bibitem{DrinSok} V.~Drinfel'd, V.~V.~Sokolov.
\textit{Lie Algebras and equations of Korteweg - de Vries type.}
Sov. J. Math. {\bf 30}, 1975--2036 (1985).

\bibitem{DS} N. Dunford, J. T. Schwartz. \textit{Linear operators.
vol. 2, Spectral theory. Self-adjoint operators in Hilbert space.}
 (1963), Interscience Publishers, Inc., NY.

\bibitem{FaTa} L.~D.~Faddeev, L.~A.~Takhtadjan. \textit{Hamiltonian
methods in the theory of solitons}. (Springer Verlag, Berlin, 1987).


\bibitem{FlaNewRat} H. Flaschka, A. C. Newell, T. Ratiu.
\textit{Kac-Moody Lie algebras and soliton equations. II. Lax equations
associated with $A_1^{(1)}$.}  Physica D {\bf 9D}, 300-323 (1983).

\bibitem{FokAnd} A. Fokas, R. L. Andersson.
\textit{On the use of isospectral eigenvalue problems for
obtaining hereditary symmetries for Hamiltonian systems.}
J. Math. Phys. {\bf 23}, No~6, 1066-1073 (1982).

\bibitem{Gakhov} F. D. Gakhov. \textit{Boundary value problems.} Translated
from Russian ed. I. N. Sneddon, (Oxford, Pergamon Press, 1966).

\bibitem{GelDick} I. M. Gel'fand, L. A. Dickey. Funct. Anal. Appl. {\bf
11}(2), 11 (1977) (In Russian).

\bibitem{Ge2} V.~S.~Gerdjikov.
\textit{On the spectral theory of the  integro-differential
operator $\Lambda$, generating  nonlinear  evolution equations.  }
Lett. Math. Phys. {\bf 6,} n.~6, 315--324, (1982).

\bibitem{Ge} V.~S.~Gerdjikov.
\textit{ Generalized Fourier transforms for the soliton equations. Gauge
covariant formulation.} Inverse Problems {\bf 2}, n.~1, 51--74, 1986.\\
--- \textit{Generating operators for the nolinear evolution
   equations of soliton type related to  the  semisimple Lie algebras.}
Doctor of Sciences Thesis,  1987, JINR, Dubna, USSR,  (In Russian).

\bibitem{Ge3} V.~S.~Gerdjikov.
\textit{  $Z_N$--reductions and  new  integrable  versions  of
    derivative nonlinear Schr\"o\-dinger equations.}
    In Nonlinear  evolution equations: integrability  and  spectral
    methods, Ed.~A.~P.~Fordy, A.~Degasperis, M.~Lakshmanan, Manchester
    University Press, (1981), p.~367--379.

\bibitem{Ge4} V.~S.~Gerdjikov.
{\it    Generalised Fourier transforms for  the  soliton
     equations. Gauge covariant formulation. }
     Inverse Problems {\bf 2,} n.~1, 51--74, (1986).
\bibitem{Ge5}  V.~S.~Gerdjikov.
\textit{Complete integrability, gauge equivalence and Lax representations
of the inhomogeneous nonlinear evolution equations.}
  Theor. Math. Phys.   {\bf 92},  374--386 (1992).

\bibitem{GGKI} V. S. Gerdjikov, G. G. Grahovski, R. I. Ivanov and N. A.
Kostov. \textit{$N $-wave interactions related to simple Lie algebras.\\
--- $\bbbz_2$- reductions and soliton solutions}. Inverse Problems
{\bf 17}, 999-1015 (2001).

\bibitem{GeIv} V.~S.~Gerdjikov, M.~I.~Ivanov.  {\it  Expansions
over the ``squared'' solutions and  the inhomogeneous nonlinear
     Schr\"o\-dinger equation.} Inverse Problems {\bf 8}, 831--847 (1992).

\bibitem{GeKh} V. S. Gerdjikov, E. Kh. Khristov.
{\it    On the evolution equations solvable with the inverse scattering
problem. I. The spectral theory. }
Bulgarian J. Phys. {\bf 7,} No.1, 28--41, (1980).  (In Russian);\\
--- \textit{    On the evolution equations solvable with the inverse scattering
problem. II. Hamiltonian structures and Backlund transformations.}
Bulgarian J. Phys. {\bf 7,} No.2, 119--133,  (1980)  (In Russian).

\bibitem{GeKu1} V.~S.~Gerdjikov, P.~P.~Kulish.
\textit{Complete integrable Hamiltonian systems related to the
non--self--adjoint Dirac operator.}
Bulgarian J. Phys. {\bf 5,} No.4,  337--349, (1978),  (In Russian).

\bibitem{GeKu}   V.~S.~Gerdjikov, P.~P.~Kulish.
 \textit{ The generating operator for the $n\times n$ linear system.}
Physica D, {\bf 3D}, n.~3, 549--564, 1981.

\bibitem{GeYa*85}   V.~S.~Gerdjikov, A.~B.~Yanovsky.
\textit{Gauge covariant formulation of the generating operator. II.
Systems on homogeneous spaces.} Phys.\ Lett.\ A, {\bf 110A}, n.~1, 53--58,
1985.

\bibitem{GeYa*86}   V.~S.~Gerdjikov, A.~B.~Yanovsky.
\textit{Gauge covariant formulation of the generating operator. I.}
Commun.\ Math.\ Phys.\, {\bf 103A}, n.~4, 549--568, 1986.

\bibitem{GeYa} V.~S.~Gerdjikov, A.~B.~Yanovski. \textit{Completeness of
the eigenfunctions for the Caudrey--Beals--Coifman system.}
J. Math. Phys. {\bf 35}, no.~7, 3687--3725 (1994).

\bibitem{Helg}  S.~Helgasson.
\textit{Differential geometry, Lie groups and symmetric spaces.}
   Academic Press, 1978.

\bibitem{JiMi}
   M.~Jimbo, T.~Miwa.
   \textit{Solitons and infinite dimensional algebras.}
   Publications RIMS {\bf 19}, 943--1000, 1983.

\bibitem{Kac} V. G. Kac. \textit{Infinite dimensional Lie algebras.}
Progress in Mathematics, vol. {\bf 44}, Boston, Birkhauser,
1983.\\
V.G.Kac, A. K. Raina. \textit{Bombay lectures on highest
weight representations of infinite dimensional Lie algebras.}
Advanced series in Math. Phys. vol. {\bf 2}, (1987).

\bibitem{Kau*76}     D.~J.~Kaup.
   \textit{Closure of the squared Zakharov--Shabat eigenstates.}
   J.\ Math.\ Annal. Appl.\ {\bf 54}, n.~3, 849--864, 1976.

\bibitem{Kaup} D. J. Kaup. \textit{The three-wave interaction -- a non-dispersive phenomenon.}
Studies in Appl. Math. {\bf 55}, 9-44 (1976); \\
D. J. Kaup, A. Reiman, A. Bers. Rev. Mod. Phys. \textit{Space-time
evolution of nonlinear three-wave interactions. I. Interaction in
a homogeneous medium. } {\bf 51}, 275-310 (1979).

\bibitem{KauNew}   D.~J.~Kaup, A.~C.~Newell.
 \textit{  Soliton equations, singular dispersion relations and moving
eigenvalues.}   Adv.\ Math.\ {\bf 31}, 67--100, 1979.

\bibitem{Konop} B. G. Konopelchenko
\textit{Solitons in Multidimensions.  Inverse Spectral Transform Method.}
World Scientific, Singapore, 1993.

\bibitem{PPK} P. P. Kulish. \textit{ Quantum difference nonlinear Schrodinger equation.}
 Lett.\ Math.\ Phys.\ {\bf 5}, 191--197, 1981.

\bibitem{KuRe}    P. P. Kulish, A. G. Reiman
\textit{Hamiltonian structure of polynomial bundles.}
    {\it Sci. Notes. LOMI seminars} {\bf 123} 67 - 76, (1983) (In
Russian); Translated in J. Sov. Math. {\bf 28}, 505-513 (1985);\\
M. A. Semenov-Tyan-Shanskii. \textit{Classical $r $-matrices and
the method of orbits.}  { Sci. Notes. LOMI seminars} {\bf 123} 77
- 91, (1983) (In Russian); Translated in J. Sov. Math. {\bf 28},
513-523 (1985).

\bibitem{Man} S. V. Manakov. \textit{An example of a completely integrable nonlinear
wave field with non-trivial dynamics (Lee model).} Teor. Mat.
Phys. {\bf 28}, 172-179 (1976).

\bibitem{Mikh} A. V. Mikhailov
\textit{ The reduction problem and the inverse scattering problem.}
Physica D, {\bf 3D}, n.~1/2, 73--117, 1981.

\bibitem{Miu} R. Miura, (editor).  \textit{B\"acklund transformations.}
Lecture Notes in Math., vol. {\bf 515}, Berlin, Springer (1979).

\bibitem{ReySTSh*80b} A. G. Reymann, M. A. Semenov-Tian Shanski.
\textit{The jets  algebra and nonlinear partial differential equations.}
 DAN USSR  (Reports of the USSR Academy), {\bf 251}, No 6, p.1310-1314,
(1980) (In Russian).

\bibitem{SegWil} G. Segal, G. Wuilson.
\textit{Loop groups and equations of KdV type.} Publ. IHES, vol. {\bf 61},
5-65 (1985).

\bibitem{Sha} A. B. Shabat. \textit{The inverse scattering problem for a
  system of differential equations.}
{}Functional Annal. \& Appl. {\bf 9}, n.3, 75 (1975) (In Russian);\\
--- \textit{The inverse scattering problem.}
Diff.  Equations {\bf 15}, 1824 (1979) (In Russian).

\bibitem{M24} E. C. Titchmarsch.
\textit{Eigenfunctions expansions associated with second  order
differential equations. Part I. Expansions over the eigenfunctions of
$(d/dx)^2 + \lambda -q(x) $.} (Oxford, Clarendon Press, 1958).

\bibitem{M8}   N.~P.~Vekua.
\textit{Systems of singular integral equations.} Translated from Russian
by A.  G.  Gibs and G. M. Simmons, (Gr\"oningen, P. Noordhoff Ltd., The
Netherlands, 1967).

\bibitem{ZMNP}  V.~E.~Zakharov, S.~V.~Manakov, S.~P.~Novikov,
L.~I.~Pitaevskii. \textit{Theory of solitons: the inverse scattering
method.}\, (Plenum, N.Y.: Consultants Bureau, 1984).

\bibitem{ZaMa} V.~E.~Zakharov, S.~V.~Manakov.  \textit{The theory of
resonant interaction of wave packets in nonlinear media.} Sov.  Phys.
JETP {\bf 69}, 1654 (1975) (In Russian).

\bibitem{ZaMa133} V.~E.~Zakharov, S.~V.~Manakov.
\textit{Multidimensional nonlinear integrable systems and methods for
constructing their solutions.} {\it Sci. Notes. LOMI seminars} {\bf 133}
77 - 91, (1984) (In Russian); Translated in J. Sov. Math. {\bf 31},
3307-3316 (1985).

\bibitem{ZaMi} V.~E.~Zakharov, A.~V.~Mikhailov.
   \textit{On the  integrability of classical spinor models in
two--dimensional space--time.} Commun.\ Math.\ Phys.\ {\bf 74}, n.~1,
   21--40, 1980;\\ ---  \textit{Relativistically invariant two-dimensional
   models of field theory which are integrable by means of the inverse
   scattering problem method.} Zh.~Eksp. Teor. Fiz. {\bf 74} 1953, (1978).


\bibitem{ZaSha} V.~E.~Zakharov, A.~B.~Shabat. \textit{A scheme for integrating nonlinear
equations of mathematical physics by the method of the inverse
scattering transform. I.} Funct. Annal. and Appl. {\bf 8}, no.~3, 43--53 (1974);\\
--- \textit{A scheme for integrating nonlinear equations of
mathematical physics by the method of the inverse scattering
transform. II.}  { Funct. Anal. Appl.} {\bf 13}(3) 13-23, (1979).

\bibitem{ZaSh*72} V.~E.~Zakharov, A.~B.~Shabat.  \textit{Exact theory of two-dimensional
self-focusing and one-dimensional self-modulation of waves in
nonlinear media.} Sov. Phys. JETP {\bf 34}, 62 (1972).

\bibitem{Zhang*98} S. Zhang.
\textit{Classical Yang-Baxter equation and low-dimensional
triangular Lie bialgebras.} Phys. Lett. A {\bf 246} 71--81 (1998).


\end{thebibliography}

\end{document}